\documentclass[runningheads,a4paper,11pt,envcountsame]{llncs}

\usepackage[a4paper,margin=2.5cm]{geometry}
\usepackage{amsmath}
\usepackage{amssymb}
\usepackage{bbding}
\usepackage[T1]{fontenc}
\usepackage{lmodern}
\usepackage{graphicx} 
\usepackage{nicefrac}
\usepackage{hyperref}
\usepackage{orcidlink}
\renewcommand{\orcidID}[1]{\orcidlink{#1}}
\usepackage{cleveref}
\usepackage{xspace}
\usepackage{xcolor}
\usepackage{tabularray}
\usepackage{standalone}
\usepackage{verbatim}
\usepackage{subcaption}

\usepackage{listings, caption}
\usepackage{algorithm}
\usepackage[noend]{algpseudocode}
\usepackage{etoolbox}

\usepackage{soul} 
\soulregister\cite7
\usepackage[commandnameprefix=ifneeded,final]{changes}
\definechangesauthor[color=magenta]{mk}
\definechangesauthor[color=orange]{yan}
\definechangesauthor[color=blue]{ir}
\definechangesauthor[color=red]{cris}
\definechangesauthor[color=red]{pedro}
\definechangesauthor[color=purple]{pierre}

\sethighlightmarkup{\IfIsColored{{\sethlcolor{authorcolor!30}\hl{#1}}}{#1}}

\usepackage[american]{circuitikz}
\usetikzlibrary{fit, backgrounds, positioning}

\pgfdeclarelayer{background}
\pgfdeclarelayer{nodelayer}
\pgfdeclarelayer{edgelayer}
\pgfsetlayers{background,edgelayer,nodelayer,main}
\tikzstyle{onegate}=[fill=white, draw=black, shape=circle]
\tikzstyle{dashed circle}=[fill=white, draw=black, shape=circle, dashed]
\tikzstyle{dashed edge}=[-, dashed]
\tikzstyle{hollow}=[fill=white, draw=black, shape=circle, inner sep=1pt]
\tikzstyle{none}=[inner sep=0mm]
\tikzstyle{graph region}=[draw, rounded corners=8pt, inner xsep=0pt, inner ysep=-3pt, fill={rgb,255: red,191; green,191; blue,191}]
\tikzstyle{smallblack}=[fill=black, draw=black, shape=circle, inner sep=1pt]
\tikzstyle{thin edge}=[-, ultra thin]
\tikzstyle{dotted circle}=[fill=white, draw=black, shape=circle, densely dotted, inner sep=2pt]

\newcommand{\x}{\times}
\newcommand{\crew}{\textsf{CREW}\xspace}
\newcommand{\pram}{\textsf{PRAM}\xspace}
\newcommand{\ram}{\textsf{RAM}\xspace}
\newcommand{\cA}{\ensuremath{\mathcal{A}}\xspace}
\newcommand{\cB}{\ensuremath{\mathcal{B}}\xspace}
\newcommand{\cO}{\mathcal{O}}
\newcommand{\cG}{\ensuremath{\mathcal{G}}\xspace}
\newcommand{\Ot}{\widetilde{\mathcal{O}}}
\newcommand{\false}{\mathit{false}}
\newcommand{\true}{\mathit{true}}
\newcommand{\diamo}{\scalebox{0.9}{$\Diamond$}}
\newcommand{\polylog}{\operatorname{polylog}}

\newcommand{\CONGEST}{\textsf{CONGEST}\xspace}
\newcommand{\CONGESTB}{\ensuremath{\textsf{CONGEST}(B)}\xspace}
\newcommand{\local}{\textsf{LOCAL}\xspace}
\newcommand{\poly}{\mathrm{poly}\xspace}
\newcommand{\complexityClass}[1]{\ensuremath{\mathrm{#1}}\xspace}
\newcommand{\NC}{\complexityClass{NC}}
\renewcommand{\P}{\complexityClass{P}}

\newcommand{\MCVP}{\textup{\textsc{M2CVP}}\xspace}
\newcommand{\CVP}{\textup{\textsc{CVP}}\xspace}

\newcommand{\LFMIS}{\textup{\textsc{LexMIS}}\xspace}
\NewDocumentCommand{\kCore}{ O{k} }{\ensuremath{#1}\textup{\textsc{-Core}}\xspace}
\newcommand{\LFDVC}{\textup{\textsc{LexCol}}\xspace}
\newcommand{\MaxFlow}{\textup{\textsc{MaxFlow}}\xspace}
\newcommand{\Disj}[3]{\ensuremath{\textup{\textsc{Disj}}^{#1}(#2,#3)}\xspace}
\newcommand{\Degen}{\textup{\textsc{Degeneracy}}\xspace}

\newcommand{\CoreValue}{\textup{\textsc{CoreValue}}\xspace}
\NewDocumentCommand{\kDeg}{ O{k} }{\ensuremath{#1}\textup{\textsc{-Degenerate}}\xspace}

\newcommand{\MovingCut}{\textup{\textsc{MovingCut}}}
\NewDocumentCommand{\DegenApprox}{ O{(2-\varepsilon)} }{\ensuremath{#1}\textup{\textsc{-ApproxDegeneracy}}\xspace}

\newcommand{\CC}{\mathrm{CC}}
\NewDocumentCommand{\HistPC}{O{\ell} O{m}}{\ensuremath{\mathrm{HistPC}_{#1}^{#2}}\xspace}
\newcommand{\zo}{\{0,1\}}
\newcommand{\Fin}[1]{F^{\downarrow}_{#1}}
\newcommand{\Fout}[1]{F^{\uparrow}_{#1}}
\newcommand{\Tin}[1]{T^{\downarrow}_{#1}}
\newcommand{\Tout}[1]{T^{\uparrow}_{#1}}
\newcommand{\fin}[2]{f^{\downarrow}_{#1 #2}}
\newcommand{\fout}[2]{f^{\uparrow}_{#1 #2}}
\newcommand{\tin}[2]{t^{\downarrow}_{#1 #2}}
\newcommand{\tout}[2]{t^{\uparrow}_{#1 #2}}
\newcommand{\lgm}{\lceil \log m \rceil}
\newcommand{\Bin}[2]{B^{\downarrow}_{#1}({#2})}
\newcommand{\Bout}[2]{B^{\uparrow}_{#1}({#2})}
\newcommand{\xx}[2]{x_{#1 #2}}
\newcommand{\yy}[2]{y_{#1 #2}}

\newcommand{\prnt}{\mathrm{p}}

\newcommand{\mkup}{\mathrm{up}}
\newcommand{\mkdown}{\mathrm{down}}

\spnewtheorem*{definition*}{Definition}{\bfseries}{\rmfamily}
\spnewtheorem*{remark*}{Remark}{\itshape}{\rmfamily}
\spnewtheorem{observation}[theorem]{Observation}{\bfseries}{\itshape}
\crefname{observation}{Observation}{Observations}
\Crefname{observation}{Observation}{Observations}

\providecommand{\qedsymbol}{\squareforqed}
\def\qed{\ifmmode\qedsymbol\else{\unskip\nobreak\hfil
\penalty50\hskip1em\null\nobreak\hfil\qedsymbol
\parfillskip=0pt\finalhyphendemerits=0\endgraf}\fi}
\let\origendproof\endproof
\renewcommand{\endproof}{\qed\origendproof}

\let\subparagraph\paragraph

\title{Lower Bounds for the Distributed Computation of Graph Degeneracy and \texorpdfstring{$k$}{k}-Core Decomposition\texorpdfstring{\thanks{This work was carried out within the framework of the STIC-AmSud ECO-DIST project (24-STIC-06) between Brazil, Chile, and France, supported by ANID (Chile, grant AMSUD240005) and CAPES (Brazil, grant 88881.985548/2024-01; Finance Code 001).}}{}}

\titlerunning{Lower Bounds for the Distributed Computation of Degeneracy and $k$-Core}

\author{Lélia Blin\orcidID{0000-0003-0342-9243}\inst{1} \and
Yan S.\ Couto\orcidID{0009-0005-5850-8696}\inst{2}\thanks{Additional support from the São Paulo Research Foundation (FAPESP-Brazil), Process Number 2024/18049-0.} \and
Cristina G.\ Fernandes\orcidID{0000-0002-5259-2859}\inst{2}\thanks{Additional support from FAPESP-Brazil, Process Number 2023/03167-5, and 
from the National Council for Scientific and Technological Development (CNPq-Brazil), Process Numbers 404315/2023-2 and 312511/2025-6.} \and
Pierre Fraigniaud\orcidID{0000-0003-4534-4803}\inst{1}\thanks{Additional support from ANR projects ENEDISC (ANR-24-CE48-7768-01) and PREDICTIONS (ANR-23-CE48-0010).} \and \\
Marcos Kiwi\orcidID{0000-0003-4171-2656}\inst{3}\thanks{Additional support from FB210005, BASAL funds for centers of excellence from ANID-Chile.} \and
Pedro Montealegre\orcidID{0000-0002-2508-5907}\inst{4}\thanks{Additional support from FONDECYT 1230599, ANID-MILENIO-NCN2024\_103 and from FB210005, BASAL funds for centers of excellence from ANID-Chile.} \and
Ivan Rapaport\orcidID{0000-0002-2969-5083}\inst{3}\thanks{Additional support from FB210005, BASAL funds for centers of excellence from ANID-Chile, and FONDECYT 1261209.}}

\authorrunning{L. Blin, Y. Couto, C. Fernandes, P. Fraigniaud, M. Kiwi, P. Montealegre, I. Rapaport}

\institute{Institut de Recherche en Informatique Fondamentale, Université Paris Cité and CNRS, France\\
\email{lelia.blin@irif.fr}, \email{pierre.fraigniaud@irif.fr} \and
Departamento de Ciência da Computação, Universidade de São Paulo, Brazil\\
\email{yancouto@ime.usp.br}, \email{cris@ime.usp.br} \and
Departamento de Ingeniería Matemática and CMM, Universidad de Chile, Chile\\
\email{mk@dim.uchile.cl}, \email{rapaport@dim.uchile.cl} \and
Facultad de Ingeniería y Ciencias, Universidad Adolfo Ibáñez, Chile\\
\email{p.montealegre@uai.cl}}

\begin{document}

\maketitle

\begin{abstract}
    The degeneracy of a graph $G$ is the maximum integer $k$ such that $G$ has a non-empty $k$-core, where the $k$-core is the largest induced subgraph with minimum degree~$k$. Computing the $k$-core or the degeneracy of a graph seems to be inherently sequential, and indeed Anderson and Mayr (Stanford Tech.\ Rep., 1984) have shown that computing the $k$-core of a graph~$G$ is P-complete. In the decentralized  setting, Chan, Mauro, and Sun (IPDPS 2019) have shown that the sequential peeling algorithm can be distributed so that the core value
    of every node, i.e., the largest integer $k$ for which the node belongs to the $k$-core, is computed in $\cO(n)$ rounds in $n$-node networks, where adjacent nodes exchange only $\cO(\log n)$ bits at each round, i.e., in the standard \CONGEST model. However, while $\Omega(D)$ rounds is a trivial lower bound for these problems in networks of diameter~$D$, this does not rule out the existence of an algorithm running in $\cO(D)+o(n)$ rounds in \CONGEST, as it is the case for many very important graph problems, including MST. We show that it is unlikely that such an algorithm exists, or, if it does exist, then it requires $\widetilde{\Omega}(\sqrt{n})$ rounds.

    Specifically, our first main result states that, for any $\varepsilon>0$,  approximating the degeneracy up to a multiplicative factor $2-\varepsilon$ requires $\widetilde{\Omega}(\sqrt{n})$ rounds in \CONGEST even in networks of diameter $\cO(\log n)$, and still $\widetilde{\Omega}(\sqrt[3]{n})$ rounds in  networks of constant diameter. These bounds hold even for randomized Monte-Carlo algorithms. They are obtained using the lower bound techniques developed by Das Sarma et al.\ (STOC 2011) revisited for capturing degeneracy. We also show that improving this lower bound seems challenging, as we demonstrate that the newer frameworks by Abboud et al. (ACM Trans.\ Alg., 2021), and by Haeupler, Wajc, and Zuzic (STOC 2021)
    cannot be used as black boxes to obtain a better lower bound for degeneracy.

    Our second main result provides a better bound, albeit conditional. It states that, unless the well-known \emph{circuit value problem} (CVP) admits a parallel polynomial speedup ---~an outcome considered unlikely~--- it holds that, for every~$\varepsilon>\delta>0$, deciding whether a graph is 2-degenerate cannot be solved in $\cO(D+ n^{1-\varepsilon})$ rounds with $\cO(n^{\delta})$-time parallel local computation per round at each node. In other words, unless CVP admits a parallel polynomial speedup, getting a truly sublinear round-complexity in \CONGEST for degeneracy would require to exploit the unlimited local computation power of \CONGEST way beyond what is usually done (most algorithms in \CONGEST can actually be implemented in $\textsf{LOGSPACE}$ at each node).

\end{abstract}

\setcounter{tocdepth}{3}
\tableofcontents
\newpage
\setcounter{page}{1}
\section{Introduction}
\label{sec:introduction}

\subsection{General Objective}

Finding the degeneracy of a graph is a subproblem of the $k$-core decomposition problem. Recall that a graph~$G$ is called~\emph{$d$-degenerate} if every subgraph of~$G$ has a vertex of degree at most~$d$. The minimum~$d$ for which~$G$ is~$d$-degenerate is called the~\emph{degeneracy} of~$G$, denoted by~$d(G)$. Degeneracy is a classical measure of graph sparsity, and it is within a constant factor from other important sparsity measures, including \emph{arboricity}, \emph{expansion}, etc. In particular, for every graph $G$, if $d$ denotes its degeneracy, and $a$ its arboricity, then $a \leq d \leq 2a-1$.
Both arboricity and degeneracy play a central role in the design of distributed graph algorithms, especially for subgraph detection~\cite{korhonen_deterministic_2018,nikabadi_beyond_2022}, coloring \cite{barenboim2013distributed,ghaffari2017simple,kothapalli2011distributed}, dominating set \cite{dory2022near} and matching \cite{czygrinow2009fast}.

Also recall that the~\emph{$k$-core} of~$G$ is the largest induced subgraph of~$G$ that has minimum degree at least~$k$. Note that the $k$-core is well-defined and unique as, for any two sets of nodes $U\subseteq V$ and $U'\subseteq V$, if the induced subgraphs $G[U]$ and $G[U']$ satisfy the degree condition, then so does $G[U\cup U']$. If~$G$ has a nonempty~$k$-core, then~$G$ is not~$(k-1)$-degenerate, that is,~$d(G) \geq k$. More precisely, the degeneracy of~$G$ is the maximum~$k$ for which it has a nonempty~$k$-core.
Given a node $v$ of a graph~$G$, the \emph{core value} of $v$ is the largest~$k$ such that $v$ belongs to the $k$-core of~$G$.
The concept of $k$-core was introduced by Seidman~\cite{seidman1983network} as a measure of social cohesion. The core value of a node can be interpreted as its level of centrality in the network, reflecting how much it is integrated within the network, as well as a measure of robustness against cascading failures~\cite{BhawalkarKLRS12,Carmi2007}.
The applications of core values are diverse and well studied, spanning a wide range of domains, including community detection~\cite{giatsidis2011evaluating,matula1983smallest}, virus propagation~\cite{kitsak2010identification}, data pruning~\cite{bader2003automated}, graph visualization~\cite{alvarez2005large}, as well as various areas of the natural sciences such as
biology~\cite{emerson2015k},
ecology~\cite{morone2019k},
neuroscience~\cite{wood2015minimal},
and physics~\cite{morone2019jamming}, among others. For a survey,
see~\cite{malliaros2020core}.

In this paper, we study the distributed complexity of these two important problems, i.e., of determining (or approximating) the degeneracy of a network~$G$, and of computing the core values of its nodes. More specifically, we consider the following problems.

\begin{definition}
    \label{def:definitions-of-our-problems} ~
    \begin{itemize}
        \item The \Degen problem asks all nodes to output the degeneracy $d(G)$ of the network.

        \item For $\alpha>1$, the $\DegenApprox[\alpha]$ problem asks each node to output an~$\alpha$-approxima\-tion of the degeneracy, that is, every node $v$ must output a value~$c_v$ such that~${d(G) \leq c_v \leq \alpha \cdot d(G)}$.

        \item For $k\geq 0$, the \kDeg problem asks all nodes to output ``accept'' if the network is $k$-degenerate, and at least one node must output ``reject'' otherwise.

        \item For $k\geq 0$, the \kCore problem asks each node to output ``yes'' if it is in the $k$-core of the network, and ``no'' otherwise.

        \item The \CoreValue problem asks each node to output the largest integer $k$ for which it belongs to the $k$-core of the network.
    \end{itemize}
\end{definition}

We tackle these problems in the standard \CONGEST model of distributed computing in networks~\cite{Peleg2000}.
Recall that the \CONGEST model is a synchronous, message-passing  model used to analyze distributed graph algorithms. It views any (connected) graph as a network in which every node is a computing entity (e.g., a Turing machine), and where the nodes exchange messages along the edges of the graph, viewed as communication links. In an $n$-node network, nodes are given distinct identifiers (IDs) in a polynomial range, i.e., each ID consumes $\cO(\log n)$ bits to be stored. Computation is performed as a sequence of synchronous \emph{rounds}.  Specifically, all nodes of an $n$-node network start at the same time and, at each round, every node (1)~sends an~$\cO(\log n)$-bit message to each of its neighbors, (2)~receives the messages sent to it by its neighbors, and (3)~performs some local computation.

\subsection{The Quest for Sublinear Algorithms}

In \CONGEST, the network is subject to two major limitations, namely \emph{locality} and \emph{bandwidth}. Indeed, a piece of information emitted by a node~$v$ at round $t$ may only be known, at round $t+r$, to the nodes in the radius-$r$ ball around~$v$ in the network, and a node can send and receive~$\cO(\log n)$ bits per round through each of its incident edges. As a consequence, even ``local problems'' such as detecting the presence of a 4-cycle can require $\widetilde{\Omega}(\sqrt{n}\,)$ rounds\footnote{The $\widetilde{\cO}$, $\widetilde{\Omega}$, and $\tilde{o}$ notation hides polylogarithmic factors.}~\cite{DruckerKO13}, and ``global problems'' such as computing a minimum-weight spanning tree (MST) require $\Omega(D)$ rounds on networks of diameter~$D$~\cite{Peleg2000}.
Nevertheless, many networks coming from practical situations have sublinear diameter, so \CONGEST algorithms running in $o(n)$ rounds on $n$-node networks with sublinear diameter are particularly interesting.
Indeed, many problems admit such algorithms, including the aforementioned MST problem, which can be solved in \CONGEST in $\cO(D+\sqrt{n}\log^\star n)$ rounds~\cite{GarayKP98,KuttenP98}. Our main goal is to figure out whether computing degeneracy or $k$-core decomposition also admits \CONGEST algorithms running in $o(n)$ rounds on networks with sublinear diameter, or whether they are ``hard'' in the sense that they require $\Omega(n)$ rounds, regardless of the diameter.

Solving \kDeg in \CONGEST suffers from the same limitation as when computing an MST, even for $k=1$. Indeed, solving~\kDeg[1] requires $\Omega(D)$ rounds on networks of diameter~$D$, merely because any distributed  algorithm for \kDeg[1] has to distinguish between a path and a cycle. (A path has degeneracy~1 while a cycle has degeneracy~2.)
On the other hand, the sequential \emph{peeling} algorithm~\cite{batagelj_m_2003}, which peels off a smallest degree node at every step, computes the core value of each node in sequential time $\cO(n+m)$ on $n$-node graphs with $m$ edges, in particular finding the degeneracy of the graph.
As we will see, the peeling algorithm can easily be implemented in \CONGEST, to run in~$\cO(n)$ rounds on $n$-node networks.
The main goal of this paper is to close the gap between these two bounds $\Omega(D)$ and $\cO(n)$. The central question addressed in this paper is therefore the following:
\begin{center}
    \fbox{
        \begin{minipage}{.9\textwidth}
            \sl Do there exist \CONGEST algorithms solving any of the problems in Definition~\ref{def:definitions-of-our-problems} in $\cO(D)+o(n)$ rounds on $n$-node networks with diameter~$D$?
        \end{minipage}
    }
\end{center}

The problems in Definition~\ref{def:definitions-of-our-problems} can be partially ordered as a function of their complexities in \CONGEST. For instance, an algorithm  solving \Degen also solves \kDeg (if~$d(G)=d$ then all nodes accept whenever $k\geq d$, and all nodes reject otherwise). Similarly, an algorithm solving~$\kCore[(k+1)]$ also solves \kDeg  (every node outputting no accepts, and every node outputting yes rejects). Moreover, an algorithm solving \CoreValue can be used to solve both \Degen, and \kCore. For \kCore, the answer is direct. For \Degen, it suffices to find the maximum core value, which can be done in $\cO(D)$ additional rounds. Overall, among the problems in Definition~\ref{def:definitions-of-our-problems}, and letting aside the approximated variants, the easiest problem is \kDeg, and the hardest is \CoreValue.

\subsection{Our Results}

We provide upper and lower bounds for the \CONGEST complexities of the problems in Definition~\ref{def:definitions-of-our-problems}, and for related problems. We mostly provide lower bounds, except for one special case deserving specific attention.

\subsubsection{Upper Bound}

The sequential peeling algorithm can be implemented in \CONGEST to solve \CoreValue (and therefore \Degen, \kCore, and \kDeg as well) in $\cO(n)$ rounds on $n$-node networks --- see~\cite{chan_distributed_2019}.
The resulting distributed algorithm may however take~$\Omega(n)$ rounds even on constant-diameter networks.
Nevertheless, we show that solving \kCore[2], and therefore deciding \kDeg[1] as well, can actually be solved in $\cO(D)$ rounds. This will contrast with our lower bounds for \kDeg when $k\geq 2$.
Note that the dependence on $D$ for \kCore[2] is optimal in the sense that any algorithm solving \kCore[2] must run in $\Omega(D)$ rounds.

Our algorithm relies on the fact that a node $v$ belongs to the $2$-core if and only if it belongs to a cycle, or to a path whose endpoints belong to a cycle.
This characterization allows us to implement an algorithm on top of a BFS tree through an upcast phase followed by a downcast phase. Local information is aggregated toward the root, and the resulting global information is then disseminated back to all nodes. Since both phases operate along a BFS tree, the overall number of rounds is bounded by $\cO(D)$.

\subsubsection{Lower Bounds}
\label{subsubsec:our-results-lower-bounds}

The lower bounds are the main results in the paper. In short, we provide a series of results indicating that the existence of a \CONGEST algorithm for computing degeneracy in $\tilde{o}(n)$ rounds on networks of diameter~$o(n)$ is unlikely, and that a round complexity $\widetilde{o}(\sqrt{n})$ is in fact unconditionally impossible even in networks of diameter $\cO(\log n)$. Our $\widetilde{\Omega}(n)$ lower bound for degeneracy is conditional: any sublinear-round algorithm would either imply a major breakthrough in complexity theory or require unusually expensive local computation, far beyond the computational resources ---typically, \textsf{LOGSPACE}--- sufficient for most known \CONGEST algorithms. It is in fact difficult to envision how such additional local power could overcome the communication bottleneck.

\paragraph*{Unconditional Lower Bounds}

Our first result uses the Set-Disjointness problem from communication complexity to obtain a lower bound of~$\widetilde{\Omega}(\sqrt{n})$ rounds for any \CONGEST algorithm computing the degeneracy of the network. More specifically, we show that, under the \CONGESTB model, that is, under the general version of \CONGEST in which every two neighbors can exchange $B$-bit messages at each round,  deciding \kDeg[2] on $n$-node networks of diameter $2p+8$ requires $\Omega((\frac{n}{pB})^{\nicefrac{p}{2p+1}})$ rounds.  In particular, in $\CONGEST$, deciding \kDeg[2]  requires $\widetilde{\Omega}(\sqrt{n})$ rounds even in networks of diameter $\cO(\log n)$, and $\widetilde{\Omega}(\sqrt[3]{n})$ rounds in networks of constant diameter.

These lower bounds hold even for randomized Monte-Carlo algorithms that succeed with  probability~$\nicefrac23$. In fact, they also hold for the approximation variant of the problem. Specifically, we show that, for any~$\varepsilon>0$, \DegenApprox requires  $\Omega(\varepsilon^2 (\frac{n \varepsilon^2}{pB})^{\nicefrac{p}{2p+1}})$ rounds to be solved in $n$-node networks of diameter $\cO(p+\nicefrac{1}{\varepsilon})$ under the \CONGESTB model.
In particular, for every fixed $\varepsilon$, \DegenApprox requires $\widetilde{\Omega}(\sqrt{n})$ rounds in \CONGEST even in networks of diameter $\cO(\log n)$, and $\widetilde{\Omega}(\sqrt[3]{n})$ rounds in  networks of constant diameter.

We also show that the techniques developed by Haeupler, Wajc, and Zuzic~\cite{haeupler_network_2020,haeupler_universally-optimal_2021} cannot derive a lower bound for \Degen better than $\widetilde{\Omega}(\sqrt{n})$, and that the techniques developed by Abboud et al.~\cite{AbboudCKP21} cannot derive non-trivial lower bounds for \Degen. (We can however show how to extend these latter techniques to get an $\widetilde{\Omega}(\sqrt[3]{n})$-round lower bound for \Degen, by a reduction from the  \emph{pointer chasing} problem; Still, this bound is worse than our $\widetilde{\Omega}(\sqrt{n})$ lower bound.)

\paragraph*{Conditional Lower Bounds}

Our unconditional $\widetilde{\Omega}(\sqrt{n})$ lower bound on the number of rounds demonstrates that there cannot exist super-efficient algorithms computing the degeneracy of networks (e.g., in $\mbox{polylog}(n)$ rounds). However, this bound does not prevent the existence of a sublinear  algorithm, i.e., running in $\cO(n^{1-\varepsilon})$ rounds for some $0<\varepsilon \leq \nicefrac12$. We show that it is unlikely that such an algorithm exists.

Our second lower bound result depends on the computational capacity of individual nodes, that is, on the amount of local computation they can perform in each round. Specifically, we show that, on $n$-node networks, if there exists a \CONGEST algorithm solving~\kDeg[2] in $R(n)$ rounds with~$T(n)$ parallel computation time per round at each node   (on a \pram with $\poly(n)$ processors), then~\kDeg[2] can be solved by a \pram algorithm that runs in time $\widetilde{\cO}(R(n)\cdot T(n))$ using $\poly(n)$ processors.

A direct consequence of this theorem is that \kDeg[2] requires~${\Omega(D + n^{1-\varepsilon})}$ rounds in $n$-node networks of diameter~$D$ under the \CONGEST model with~$\cO(n^{\delta})$ parallel computation time per round at every node whenever~$\varepsilon>\delta>0$, unless the \emph{circuit value problem}\footnote{The Circuit Value Problem (\CVP) is the decision problem of determining whether a given Boolean circuit evaluates to true on a given input.} (\CVP) admits a parallel polynomial speedup.
As establishing such a speedup would be a \highlight[id=cris,comment={weird that ``breakthrough'' is positive but used in a negative way}]{breakthrough} in complexity theory, our result shows that it is unlikely that a truly sublinear algorithm for degeneracy exists, or this would be at the cost of performing a huge amount of computation per round at some node. We stress the fact that our lower bound holds even in networks of constant maximum degree.

For illustrating the significance of our conditional lower bound, we note that most algorithms in \CONGEST are frugal in terms of local computation, and many can actually be implemented in $\textsf{LOGSPACE}$. In contrast, a sequential computation time $\mbox{poly}(n)$ at every node in $n$-node graphs of constant maximum degree~$\Delta$ corresponds to exponential time with respect to the amount of information $\cO(\Delta\log n)$ received at each round by each node.  Our result holds for computation power beyond $\textsf{LOGSPACE}$, up to $\cO(n^{\delta})$ parallel computation time per round at every node. As a consequence, developing efficient algorithms for~\Degen in \CONGEST would require exploiting the unlimited local computation power of this model, far beyond what is typically done.

\begin{remark*}
    Stated as is, our conditional lower bound applies only to deterministic algorithms. However, the randomized version of the conjecture, i.e., that \CVP does not admit a randomized parallel speedup, has not been disproved. Our lower bound is thus likely to hold for randomized algorithms as well.
\end{remark*}

\paragraph*{Beyond Degeneracy and $k$-Core}

Up to our knowledge, this paper is the \highlight[id=pierre,comment={Not sure of that. I will double check.}]{first to explicitly establish connections} between parallel complexity and \CONGEST lower bounds. These connections apply not only to degeneracy and $k$-core, but also to variants of these problems, as well as to other~\P-complete problems. Note that it was known~\cite{DruckerKO13} that such connections exist between the Congested Clique model and circuit complexity, which essentially hindered proving non-trivial lower bounds for the Congested Clique model.

More precisely, our conditional lower bound holds for the complexity in \CONGEST of several \P-complete graph problems, namely the ones with a reduction from \CVP satisfying some restrictive conditions. As an important illustration, this is in particular the case for maximum flow (\MaxFlow). As other examples, one can mention the resolution of problems playing an important role in \CONGEST, such as $(\Delta+1)$-coloring graphs of maximum degree~$\Delta$, or computing a maximal independent set, with the additional constraint of finding the smallest solution under some lexicographic ordering, referred to as \mbox{\LFDVC} and \mbox{\LFMIS}. Our conditional $\Omega(n)$ lower bound also applies to \MaxFlow, \LFDVC, and \LFMIS.

\subsection{Related Work}

The potential of parallelism to speed up polynomial-time sequential algorithms has been a subject of study for almost four decades~\cite{condon_theory_1994}. Roughly speaking, a sequential algorithm with running time $t(n)$ admits a \emph{polynomial parallel speedup} if there exists a parallel algorithm that uses a polynomial number of processors and runs in time $T(n)$, with $t(n)/T(n) = \Omega(n^{\varepsilon})$ for some~${\varepsilon > 0}$. Despite extensive efforts, the fundamental question of whether all problems admit a polynomial speedup remains open. Partial advances, however, shed light on this question. For instance, in the early 90s, Mak~\cite{mak1997parallelism} showed that it is always possible to achieve a polynomial speedup, albeit with an \emph{exponential} number of processors. Two decades later, Williams proved~\cite{williams2011parallelizing} that an~$\cO(\log n)$ speedup can be achieved on a log-cost \ram algorithm, when using a log-cost \pram with a \emph{polynomial} number of processors, and conjectured that this speedup might be improved to~$\cO(\log n \cdot \log \log n)$. Yet even such a modest improvement remains unproven.
Whether~\CVP admits \emph{any} polynomial parallel speedup also remains open.

There is a noticeable similarity between results on $k$-core in distributed and parallel systems.
With respect to parallel computation, Anderson and Mayr~\cite{anderson_p-complete_1984} showed that computing the~$k$-core is \P-complete,
that there exists a~$(2+\varepsilon)$-approximation for $k$-core in \NC, and that
no~$(2-\varepsilon)$-approximation in \NC exists unless~${\P=\NC}$.  In the context of distributed systems, the study of core values  was initiated in~\cite{montresor_2011}.
An $\cO(\log n)$-round
$(2 + \varepsilon)$-approximation algorithm for computing the core values in \CONGEST was designed in~\cite{chan_distributed_2019}, where it is also proved that any $(2+\varepsilon)$-approximation needs~$\Omega(\log n)$ rounds, and any~$(2-\varepsilon)$-approximation needs~$\Omega(D)$ rounds.
Our lower bounds for \kDeg[2] also hold for $(2-\varepsilon)$-approximation algorithms computing the core values, and are actually stronger. Several other works developed distributed algorithms for~$k$-core~\cite{aksu_distributed_2014,aridhi_distributed_2016,guo_experimental_2025,mandal_distributed_2017}. However, these works are based on heuristics, with a focus on algorithms that run fast on real-world instances. As we show in this paper, obtaining an efficient distributed algorithm for~$k$-core on general instances is  not feasible.

Several algorithms in \CONGEST are parameterized by degeneracy, as for instance subgraph detection~\cite{korhonen_deterministic_2018,nikabadi_beyond_2022}, and their algorithms use the existing efficient~$(2+\varepsilon)$-approximations for degeneracy. While these parameterized algorithms are efficient (in low degeneracy graphs), and while a~$(2+\varepsilon)$-approximation is enough for their use case, the dependency on the computed approximation degeneracy $d$ of the actual degeneracy may be exponential, as in the~$2^{\cO(k d)} k^k + \cO(\log n)$ rounds for finding induced tree subgraphs on~$k$ nodes~\cite{nikabadi_beyond_2022}. Our results imply that it is unfortunately not possible to lower the exponent by achieving an approximation ratio better than~2 for the degeneracy.

In the framework of streaming algorithms, a single-pass algorithm that computes a~$(1 + \varepsilon)$-approximation for every core value using $\widetilde{\cO}(n/\varepsilon^{2})$ space is presented in~\cite{esfandiari2018parallel}.  In contrast, any semi-streaming algorithm for exactly solving this problem requires  $\widetilde{\Omega}(n^{\nicefrac13})$ passes~\cite{AssadiGLMM24}.
Regarding dynamic graph algorithms,~\cite{couto_hardness_2025} established conditional lower bounds for the $k$-core problem. Namely, assuming well-established conjectures, the $k$-core problem requires $\Omega(m^{1 - \varepsilon})$ time per update or query on a graph with $m$ edges.
O'Brien and Sullivan~\cite{o2014locally} presented an algorithm that estimates
the core value of every node $v$ based on local information within a ball of bounded radius centered at~$v$. The algorithm provides an asymptotic error bound for the
core value estimation on Erd\H{o}s--R\'enyi random graphs.

\section{Preliminaries}

\subsection{\CONGEST Model}
\label{subsec:CONGEST-LOCAL}

The \CONGESTB model~\cite{Peleg2000} is a classical model of distributed computation represented by a connected simple graph\footnote{We adopt the classical graph theory notations from the book of Bondy and Murty~\cite{BM2008}.} $G=(V,E)$, where each node $v\in V$ executes the same algorithm in synchronous rounds, and the nodes collaborate to solve a graph problem defined on $G$. Every node has an identifier, which is unique in the network. It is assumed that, in an~$n$-node network, each identifier can be stored in $\cO(\log n)$ bits. Additionally, every node $v$ may receive an input. The goal is for each vertex to produce a predefined output based on the structure of the graph and the input of all nodes. In \Degen, every node must eventually output the degeneracy of the graph, while in~\kCore, it must output \emph{yes} or \emph{no} depending on whether it belongs to the $k$-core of the network.

The defining feature of the \CONGESTB model is that computation proceeds as a sequence of synchronous rounds. In each round, every node sends a message to each of its neighbors, receives messages from all of its neighbors, and performs some local computation. The size of the messages is restricted to $B$ bits.
Each node is allowed to perform an unlimited amount of local computation. Thus, the main computational cost is measured in terms of communication rounds. The standard \CONGEST model corresponds to \CONGESTB
with~$B = \cO(\log n)$, which allows each message to contain, for example, at most a constant number of node identifiers.

\begin{remark*}
    The \local model is the same as $\CONGEST(\infty)$, that is, there are no restrictions on the size of the messages exchanged. In particular, every (computable) graph problem is solvable in $\cO(D)$ rounds under the \local model in networks of diameter~$D$, by gathering the whole graph at one node (e.g., the one with minimum identifier), computing the solution locally at this node, and broadcasting that solution back to all nodes. Such a trivial solution is not possible in \CONGEST, as gathering the whole graph on a node may take $\Omega(n)$ rounds in sparse or dense graphs.
\end{remark*}

\subsection{PRAM, NC, and the Circuit Value Problem}
\label{subsec:PRAM-NC-etc}

A \emph{parallel random-access machine} (parallel \ram, or \pram for short) is a multiprocessor shared-memory abstract machine. In the so-called concurrent read exclusive write (\crew) \!\pram, at each time step, each memory register can be read by multiple processors, or be written to by a single one. There are other variants of the \pram model, which differ in how processors access the memory registers (such as \textsf{EREW}), but all of them are equivalent up to polylogarithmic factors in the running time, and polynomial factors in the number of processors.
The class \P is the set of all decision problems (or languages) decidable in
polynomial time on a (sequential) \ram.
The class \NC is the set of all decision problems (or languages) decidable in polylogarithmic time on a \pram with a polynomial number of processors.

One of the central problems that is complete for $\P$ under NC reductions is the \emph{circuit value problem} (\CVP).
A \emph{(Boolean) circuit} is a directed acyclic graph (DAG), with each node labeled from the set $\{0, 1, \textrm{AND}, \textrm{OR}, \textrm{NOT}\}$, and a specified output node. We refer to nodes and edges of the DAG as \emph{gates} and \emph{wires}, respectively. The output gate has out-degree~0, while 0- and 1-gates have in-degree~0. Gates with out-degree larger than~1 output the same value on all wires. The propagation of the values of the 0- and 1-gates throughout the AND-, OR- and NOT-gates results in a 0 or 1 value output at each gate. The \emph{circuit value} is the output value of the output gate. The \emph{size} of the circuit is its number of gates plus wires. The circuit value problem is the problem of, given a circuit, computing its value.
A circuit is \emph{monotone} if it does not have NOT-gates. We refer to \CVP restricted to monotone circuits where input gates have out-degree 1, non-input non-output gates have in- and out-degree exactly 2, and the output gate has in-degree 2, as the \emph{monotone fan-in 2 fan-out 2 circuit value problem} (\MCVP).
It is known~\cite{greenlaw_limits_1995} that \MCVP, as \CVP, is~\P-complete.

\subsection{Flows, Lexicographic MIS and \texorpdfstring{$(\Delta+1)$}{(Δ+1)}-Coloring}

Our lower bounds for degeneracy and $k$-core also apply to other problems such as, prominently,  computing \emph{maximum flows}.

\subsubsection{Maximum Flow}

Let $D = (V, A)$ be a \emph{directed} graph, let $c: A \to \mathbb{N}$ be the arc \emph{capacities}, and let $s$ and $t$ be the \emph{source} and \emph{sink} vertices, respectively. A \emph{flow} in $D$ is a mapping $f: A \to \mathbb{N}$. A flow $f$ is \emph{feasible} if $f(e) \leq c(e)$ for each arc $e \in A$, and $\sum_{uv \in A}f(uv) = \sum_{vu \in A}f(vu)$ for each~$u \in V \smallsetminus \{s,t\}$. The \MaxFlow problem asks to compute a feasible flow $f$ that maximizes $\sum_{sv \in A}f(sv)$. In \CONGEST, each node receives the orientation and capacities of its arcs, and must output the value of a (feasible and maximum) flow for each arc incident to it.

\subsubsection{Lexicographic Variants of LCL Problems}

Other problems impacted by our results include lexicographic variants of standard locally checkable labeling problems such as maximal independent set and $(\Delta+1)$-coloring. The study of such problems is standard in the context of parallel complexity~\cite{greenlaw_limits_1995}. In all these variants, we assume that every node $v$ of the graph $G = (V,E)$ is given a number $\ell(v) \in \mathbb{N}$ as input, which is assumed to be unique.
\begin{itemize}
    \item Recall that an \emph{independent set} in $G$ is a set of pairwise non-adjacent vertices of~$G$, and that an independent set is \emph{maximal} if no vertex can be added to it while maintaining the independent set property. Let $I = \{v_1, \ldots, v_r\}$ be a maximal independent set with~${\ell(v_1) < \cdots < \ell(v_r)}$. The ordered sequence~$(\ell(v_1), \ldots, \ell(v_r))$ is called the \emph{signature} of~$I$. The \LFMIS problem asks to compute the maximal independent set of~$G$ that has the smallest signature, where signatures are lexicographically ordered as words on the alphabet~$\mathbb{N}$. In \CONGEST, each node must output~1 if it is in the MIS, and~0 otherwise.

    \item A \emph{$k$-coloring} of $G$ is an assignment $c: V \to \{1,\dots,k\}$ such that $c(u) \neq c(v)$ for each~$uv \in E$. The lexicographic variant of $(\Delta+1)$-coloring is defined similarly, where $\Delta$ denotes the maximum degree of $G$.  Let $V = \{v_1, \ldots, v_n\}$ with ${\ell(v_1) < \cdots < \ell(v_n)}$. Given a~$(\Delta+1)$-coloring $c$ of~$G$, the ordered sequence~$(c(v_1), \ldots, c(v_n))$ is called the \emph{signature} of the coloring~$c$. The \LFDVC problem asks to compute the ${(\Delta+1)}$-coloring of~$G$ that has the smallest signature, when signatures are compared lexicographically as words on the alphabet~$\{1,\dots,\Delta+1\}$. In \CONGEST, each node must output its color.
\end{itemize}

Both \LFMIS and \LFDVC have sequential greedy algorithms in which one decides, node by node, in increasing order of their labels, whether they belong to the maximal independent set, or which color they should be given. We will show that there is no efficient way to parallelize or distribute this greedy algorithm.

\section{Unconditional Lower Bounds}
\label{sec:moving-cuts-lb}

We start by providing unconditional lower bounds based on the framework of ``moving cuts'' designed by Das Sarma et al.~\cite{sarma_distributed_2011}. We begin the presentation by considering the exact variant of degeneracy (deciding \kDeg[2]), and then show how to extend the bound to the approximated variant of degeneracy.

\subsection{Lower Bound for Deciding Degeneracy 2}

\begin{figure}[htb]
    \centering
    \scalebox{0.75}{\begin{tikzpicture}
	\begin{pgfonlayer}{nodelayer}
		\node [style=smallblack] (15) at (0, 0) {};
		\node [style=smallblack] (18) at (0.5, 2) {};
		\node [style=smallblack] (19) at (2.5, 2) {};
		\node [style=smallblack] (20) at (1.5, 3) {};
		\node [style=smallblack] (21) at (1, -2) {};
		\node [style=smallblack] (24) at (2, -2) {};
		\node [style=smallblack] (27) at (0, -1) {};
		\node [style=smallblack] (30) at (1, -1) {};
		\node [style=smallblack] (33) at (2, -1) {};
		\node [style=smallblack, label=left:$s$] (36) at (0, 1) {};
		\node [style=smallblack] (37) at (1, 0) {};
		\node [style=smallblack] (40) at (1, 1) {};
		\node [style=smallblack] (41) at (2, 0) {};
		\node [style=smallblack] (44) at (2, 1) {};
		\node [style=smallblack] (45) at (3, 0) {};
		\node [style=smallblack, label=right:$r$] (48) at (3, 1) {};
		\node [style=smallblack] (64) at (3, -1) {};
		\node [style=smallblack] (82) at (0, -2) {};
		\node [style=smallblack] (100) at (3, -2) {};
	\end{pgfonlayer}
	\begin{pgfonlayer}{edgelayer}
		\draw [style=thin edge] (20) to (18);
		\draw [style=thin edge] (20) to (19);
		\draw [style=thin edge] (21) to (24);
		\draw [style=thin edge] (27) to (30);
		\draw [style=thin edge] (30) to (33);
		\draw [style=thin edge, bend right, looseness=0.75] (40) to (37);
		\draw [style=thin edge, bend left, looseness=0.75] (44) to (41);
		\draw [style=thin edge] (37) to (41);
		\draw [style=thin edge, bend left, looseness=0.75] (48) to (45);
		\draw [style=thin edge] (41) to (45);
		\draw [style=thin edge, bend right, looseness=0.75] (36) to (27);
		\draw [style=thin edge, bend right, looseness=0.75] (40) to (30);
		\draw [style=thin edge, bend left, looseness=0.75] (44) to (33);
		\draw [style=thin edge] (18) to (36);
		\draw [style=thin edge, bend right, looseness=0.75] (40) to (21);
		\draw [style=thin edge] (18) to (40);
		\draw [style=thin edge, bend left, looseness=0.75] (44) to (24);
		\draw [style=thin edge] (19) to (44);
		\draw [style=thin edge] (19) to (48);
		\draw [style=thin edge, bend right, looseness=0.75] (36) to (15);
		\draw [style=thin edge] (15) to (37);
		\draw [style=thin edge] (33) to (64);
		\draw [style=thin edge, bend left, looseness=0.75] (48) to (64);
		\draw [style=thin edge] (82) to (21);
		\draw [style=thin edge, bend right, looseness=0.75] (36) to (82);
		\draw [style=thin edge] (24) to (100);
		\draw [style=thin edge, bend left, looseness=0.75] (48) to (100);
	\end{pgfonlayer}
\end{tikzpicture}}
    \caption{The graph $G(q, d, p)$ for $q = 3$, $d = 2$, and $p = 2$, that is,~$G(3,2,2)$.}
    \label{fig:ggdp-example}
\end{figure}
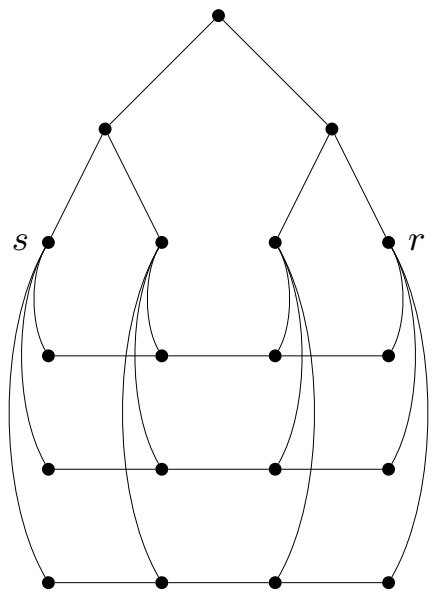

Let us define a graph family used for establishing the lower bound. For integers $p \geq 1$, $q \geq 1$, and $d \geq 2$, let $G(q, d, p)$ be the graph consisting of $q$ paths, each with $d^p$ nodes, plus a complete $d$-ary tree $T$ of depth $p$ and $d^p$ leaves. The $i$th leaf of~$T$ is connected to the $i$th vertex in each of the $q$ paths. We denote by $s$ and $r$ the first and last leaves of~$T$. See~\Cref{fig:ggdp-example} for an  illustration. The graph~$G(q, d, p)$ has~$\Theta(q\cdot d^p)$ nodes, and diameter $2p+2$.

Our lower bound in \CONGEST follows from the lower bound for the set-disjointness problem in communication complexity. Given an integer $q\geq 1$, the $q$-disjointness problem in $G(q, d, p)$, denoted by $\Disj{q}{s}{r}$, is defined as follows. Nodes $s$ and $r$ are respectively given inputs $x\in\{0,1\}^q$ and $y\in\{0,1\}^q$. The goal is for $s$ and $r$ to decide whether $x$ and $y$ are disjoint, that is, whether $\min\{x_i,y_i\}=0$ for all~$i \in \{1,\dots,q\}$. Both nodes must accept if $x$ and $y$ are disjoint, and both of them must reject otherwise.
Our lower bound is based on the following result.

\begin{lemma}[\cite{sarma_distributed_2011}]
    \label{res:disj-lower-bound}
    For any $p \geq 1$, and $d \ge 2$, $\Disj{q}{s}{r}$ on $G(q,d,p)$ with $q = d^{p+1} p B$ requires $\Omega(d^p)$ rounds to be solved in \CONGESTB, even by a randomized Monte-Carlo algorithm succeeding with probability~$\nicefrac23$.
\end{lemma}

\begin{figure}[tb]
    \centering
    \begin{tikzpicture}
	\begin{pgfonlayer}{nodelayer}
		\node [style=smallblack] (0) at (0, 1) {};
		\node [style=smallblack] (1) at (-1, 0) {};
		\node [style=smallblack] (2) at (0, -1) {};
		\node [style=smallblack] (3) at (0, 0) {};
		\node [style=smallblack] (4) at (1, 0) {};
		\node [style=dotted circle] (5) at (3, 0) {};
	\end{pgfonlayer}
	\begin{pgfonlayer}{edgelayer}
		\draw (0) to (3);
		\draw (0) to (1);
		\draw (1) to (2);
		\draw (3) to (2);
		\draw (3) to (1);
		\draw (0) to (4);
		\draw (4) to (2);
		\draw (4) to (5);
	\end{pgfonlayer}
\end{tikzpicture}
    \caption{The flyswatter gadget. If the gadget is attached to a node $v$ in $H(x,y,q,d,p)$, then the node $v$ is the node depicted by the dotted circle on the figure.}
    \label{fig:gadget-3core}
\end{figure}
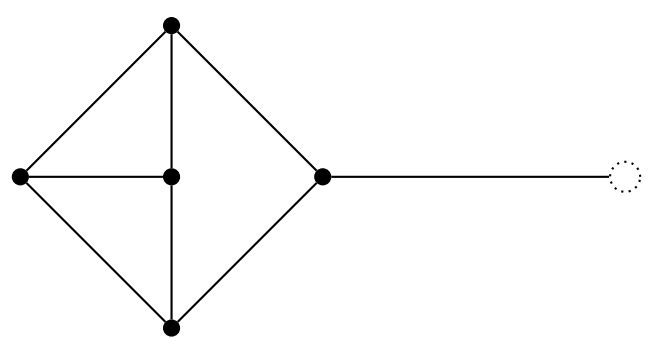

To obtain a lower bound for $\kDeg[2]$, we show that any algorithm for this problem can be used to solve $\Disj{q}{s}{r}$ on $G(q,d,p)$. This will require to modify the network $G(q,d,p)$, but we shall later argue how our modifications can be simulated by the nodes of the original $G(q,d,p)$. Our construction is based on the following graph family. For any integers $q,p \geq 1$, any binary strings $x, y \in \{0,1\}^q$, and any integer $d \geq 2$, the graph $H(x,y,q,d,p)$ is obtained by modifying~$G(q, d, p)$. The construction uses the gadget depicted on \Cref{fig:gadget-3core}, referred to as the \emph{flyswatter} gadget. Copies of this gadget are attached to vertices of $G(q,d,p)$ according to the following rules.
\begin{itemize}
    \item For every $i \in [q]$, if $x_i = 1$ then attach a flyswatter to the first node of the $i$th path.
    \item For every $i \in [q]$, if $y_i = 1$ then attach a flyswatter to the last node of the $i$th path.
    \item Attach a flyswatter to each leaf of the tree~$T$ in~$G(q, d, p)$, as well as to its root.
\end{itemize}

As the number of nodes in $H(x,y,q,d,p)$ is at most a constant factor larger than the number of nodes in $G(q,d,p)$, and as its diameter increases by at most twice the diameter of the flyswatter, the following holds:

\begin{observation}\label{res:hxydp-size}
    $H(x, y, q, d, p)$ has $\Theta(q d^p)$ nodes, and diameter $2p+8$.
\end{observation}

The way the flyswatters are attached to $G(q,d,p)$ as a function of $x$ and $y$ enables one to relate the degeneracy of $H(x,y,q,d,p)$ to the truth value of the disjointness of $x$ and $y$.

\begin{lemma}\label{res:g-prime-core}
    $H(x,y,q,d,p)$ is $2$-degenerate if and only if $x$ and $y$ are disjoint.
\end{lemma}

\begin{proof}
    Suppose first that $H=H(x,y,q,d,p)$ is~$2$-degenerate, i.e., every subgraph of $H$ has a vertex of degree at most~2. Let $H_i$ be the subgraph of $H$ containing its $i$th path, the tree $T$ of $G(q, d, p)$, and all flyswatters attached to those vertices. Note that all internal vertices of the $i$th path, all the vertices in $T$, and all vertices in the flyswatters have degree at least~3 in~$H_i$, independently of the values of $x$ and~$y$. For every $i \in \{1,\dots,q\}$, since $H$ is~$2$-degenerate, there must be a vertex of~$H_i$ with degree at most~2, which implies that at least one of the two endpoints of the $i$th path is not connected to a flyswatter. It follows that $x_i=0$ or $y_i=0$, and thus $x$ and~$y$ are disjoint.

    Let us now suppose that $H=H(x,y,q,d,p)$ is not $2$-degenerate, meaning that $H$ has a nonempty 3-core~$H'$. A flyswatter can have either all its vertices in the 3-core~$H'$, or none of them. That is, a flyswatter belongs to $H'$ if and only if it is attached to a vertex that belongs to~$H'$. Also, $H'$ cannot be completely contained in the tree~$T$. It follows that $H'$ must contain some vertex from at least one of the $q$ paths, as otherwise~$H'$ would be empty. A consequence of the fact that $H'$ contains at least one vertex $v$ from one of the paths is that $H'$ must contain all the neighbors of~$v$ because the vertices in the paths have degree at most~3 in~$H(x,y,q,d,p)$. It follows that $H'$  must actually contain a complete path, say the $i$th path, for some~$i \in \{1,\dots,q\}$. In particular, the endpoints of that path must have degree~3, which implies that they are both attached to a flyswatter. This means that $x_i=y_i=1$, and thus $x$ and $y$ are not disjoint.
\end{proof}

Now, we show how to transform an algorithm for \kDeg[2] into an algorithm for $\Disj{q}{s}{r}$, while controlling its round complexity.

\begin{lemma}\label{res:disj-to-degen}
    If there exists an algorithm $\cA$ solving \kDeg[2] on $n$-node networks of diameter $2p+8$ in $R(n)=\cO(n)$ rounds under \CONGESTB, then there exists an algorithm $\cB$ solving $\Disj{q}{s}{r}$ on $G(q,d,p)$ in $\cO(R(q\cdot d^p)+p)$ rounds under \CONGESTB. Moreover, if $\cA$ is a randomized Monte Carlo algorithm succeeding with probability at least~$\nicefrac23$, then so is~$\cB$.
\end{lemma}

\begin{proof}
    Suppose that there is an algorithm \cA for \kDeg[2] as in the statement. We describe an algorithm~$\cB$ for $\Disj{q}{s}{r}$ on $G(q, d, p)$. In the first round of~$\cB$, for each~$i \in [q]$, node~$s$ sends $x_i$ to the first node of the $i$th path, and node $r$ sends $y_i$ to the last node of the $i$th path.

    At this point, every node has enough information to determine whether it is attached to a flyswatter gadget in the construction of $H(x,y,q,d,p)$, and so can simulate the virtual execution of \cA on $H(x,y,q,d,p)$ in the actual network $G(q, d, p)$. In particular, any node~$v$ that is attached to a flyswatter gadget locally simulates the states and the exchanged messages for all (virtual) nodes in that gadget. Since the only edge leaving a flyswatter gadget connects to the node that is simulating that gadget, the simulation does not require any additional communication over the edges of~$G(q, d, p)$. From the second round, $\cB$ is thus the mere simulation of \cA on the virtual network $H(x,y,q,d,p)$.

    At the end of the simulation, the nodes output the result of~\cA, i.e., whether $H(x,y,q,d,p)$ is $2$-degenerate or not.
    More specifically, if $H(x,y,q,d,p)$ is~$2$-degenerate then all nodes accept, and if $H(x,y,q,d,p)$ is not $2$-degenerate, then at least one node rejects. To guarantee that both $r$ and $s$ reject in case $H(x,y,q,d,p)$ is not $2$-degenerate, any rejecting node floods its opinion throughout the network, in diameter rounds.
    By~\Cref{res:g-prime-core}, this algorithm $\cB$ correctly computes whether~$x$ and~$y$ are disjoint.

    As far as the round complexity of $\cB$ is concerned, note that its first round communicates at most one bit per edge, so this round is valid in~\CONGESTB for any~$B \geq 1$. The simulation in $G(q,d,p)$ of the execution of $\cA$ on $H(x, y, q,d, p)$ takes $1+R(\Theta(q\cdot d^p))$ rounds because $H(x, y, q,d, p)$ has $\Theta(q\cdot d^p)$ nodes (cf. \Cref{res:hxydp-size}).
    By \Cref{res:hxydp-size}, the final flooding consumes $\cO(p)$ additional rounds.
    Since $R(n)=\cO(n)$, we have $R(\cO(N)) = \cO(R(N))$ for any~$N$, implying that $\cB$ takes $\cO(R(q\cdot d^p)+p)$ rounds.
\end{proof}

\begin{theorem}\label{thm:kdeg2-lb}
    For any $p\geq 1$, deciding \kDeg[2] on $n$-node networks of diameter $2p+8$  requires $\Omega((\frac{n}{pB})^{\nicefrac{p}{2p+1}})$ rounds in the \CONGESTB model. This holds even for randomized Monte-Carlo algorithms succeeding with probability at least~$\nicefrac23$.
\end{theorem}

\begin{proof}
    Suppose that there is a randomized \CONGESTB algorithm for \kDeg[2] on $n$-node networks running in $R(n)$ rounds, where $R(n)=\cO(n)$. By~\Cref{res:disj-to-degen}, there is a randomized algorithm for~$\Disj{q}{s}{r}$ on the network~$G(q,d,p)$ running in~$\cO(R(q\cdot d^p))$ rounds. By~\Cref{res:disj-lower-bound}, we obtain~$R(q\cdot d^p) = \Omega(d^p)$, when~$q = d^{p+1} p B$. Since the number of nodes in the network~$H(x,y,q,d,p)$ is~$n = \Theta(q\cdot d^p) = \Theta(d^{2p+1} p B)$, and since $R(n) = \Omega(d^p)$, expressing~$d^p$ in terms of~$n$ yields~$R(n) = \Omega((\nicefrac{n}{pB})^{\nicefrac{p}{2p+1}})$, as claimed.
\end{proof}

\begin{corollary}\label{cor:kdeg2-lb}
    Deciding \kDeg[2] on $n$-node networks of diameter $\cO(\log n)$ requires $\Omega(\sqrt{\nicefrac{n}{\log^2 n}})$ rounds in \CONGEST, and deciding \kDeg[2] on $n$-node networks of diameter $10$ requires $\Omega(\sqrt[3]{\nicefrac{n}{\log n}})$ rounds in \CONGEST.
\end{corollary}

\subsection{Lower Bound for Approximate Degeneracy}

We now show that even approximating the degeneracy of a graph within a multiplicative factor smaller than~2 is hard. To do that, we introduce a graph $H_k(x,y,q,d,p)$ for every $k \geq 2$ and~$x,y \in \{0,1\}^q$. Our construction guarantees that the degeneracy of $H_k(x,y,q,d,p)$ changes by a large factor depending on whether $x$ and $y$ are disjoint, which allows us to use any $(2-\varepsilon)$-approximation for the degeneracy to solve $\Disj{q}{s}{r}$, from which a similar lower bound as the one given in~\Cref{thm:kdeg2-lb} for $\kDeg[2]$ can be obtained.

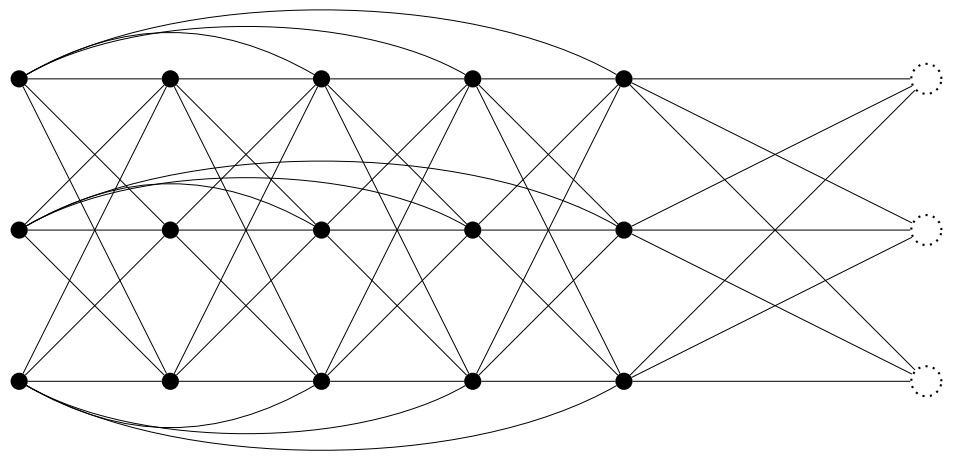
\begin{figure}[tb]
    \centering
    \begin{tikzpicture}
	\begin{pgfonlayer}{nodelayer}
		\node [style=smallblack] (0) at (0, 1) {};
		\node [style=smallblack] (6) at (1, 1) {};
		\node [style=smallblack] (12) at (2, 1) {};
		\node [style=smallblack] (18) at (3, 1) {};
		\node [style=smallblack] (21) at (4, 1) {};
		\node [style=smallblack] (24) at (0, -1) {};
		\node [style=smallblack] (25) at (1, -1) {};
		\node [style=smallblack] (26) at (2, -1) {};
		\node [style=smallblack] (27) at (3, -1) {};
		\node [style=smallblack] (28) at (4, -1) {};
		\node [style=smallblack] (29) at (0, 0) {};
		\node [style=smallblack] (30) at (1, 0) {};
		\node [style=smallblack] (31) at (2, 0) {};
		\node [style=smallblack] (32) at (3, 0) {};
		\node [style=smallblack] (33) at (4, 0) {};
		\node [style=dotted circle] (34) at (6, 1) {};
		\node [style=dotted circle] (35) at (6, 0) {};
		\node [style=dotted circle] (36) at (6, -1) {};
	\end{pgfonlayer}
	\begin{pgfonlayer}{edgelayer}
		\draw [style=thin edge] (0) to (6);
		\draw [style=thin edge] (6) to (12);
		\draw [style=thin edge] (18) to (21);
		\draw [style=thin edge] (12) to (18);
		\draw [style=thin edge, bend left] (0) to (12);
		\draw [style=thin edge, bend left, looseness=0.75] (0) to (18);
		\draw [style=thin edge, bend left, looseness=0.75] (0) to (21);
		\draw [style=thin edge] (24) to (25);
		\draw [style=thin edge] (25) to (26);
		\draw [style=thin edge] (27) to (28);
		\draw [style=thin edge] (26) to (27);
		\draw [style=thin edge, bend right] (24) to (26);
		\draw [style=thin edge, bend right, looseness=0.75] (24) to (27);
		\draw [style=thin edge, bend right, looseness=0.75] (24) to (28);
		\draw [style=thin edge] (0) to (25);
		\draw [style=thin edge] (24) to (6);
		\draw [style=thin edge] (6) to (26);
		\draw [style=thin edge] (25) to (12);
		\draw [style=thin edge] (18) to (28);
		\draw [style=thin edge] (27) to (21);
		\draw [style=thin edge] (12) to (27);
		\draw [style=thin edge] (26) to (18);
		\draw [style=thin edge] (29) to (30);
		\draw [style=thin edge] (30) to (31);
		\draw [style=thin edge] (32) to (33);
		\draw [style=thin edge] (31) to (32);
		\draw [style=thin edge, bend left] (29) to (31);
		\draw [style=thin edge, bend left, looseness=0.75] (29) to (32);
		\draw [style=thin edge, bend left, looseness=0.75] (29) to (33);
		\draw [style=thin edge] (0) to (30);
		\draw [style=thin edge] (29) to (6);
		\draw [style=thin edge] (6) to (31);
		\draw [style=thin edge] (30) to (12);
		\draw [style=thin edge] (18) to (33);
		\draw [style=thin edge] (32) to (21);
		\draw [style=thin edge] (12) to (32);
		\draw [style=thin edge] (31) to (18);
		\draw [style=thin edge] (29) to (25);
		\draw [style=thin edge] (24) to (30);
		\draw [style=thin edge] (30) to (26);
		\draw [style=thin edge] (25) to (31);
		\draw [style=thin edge] (32) to (28);
		\draw [style=thin edge] (27) to (33);
		\draw [style=thin edge] (31) to (27);
		\draw [style=thin edge] (26) to (32);
		\draw [style=thin edge] (21) to (34);
		\draw [style=thin edge] (21) to (35);
		\draw [style=thin edge] (21) to (36);
		\draw [style=thin edge] (33) to (34);
		\draw [style=thin edge] (33) to (35);
		\draw [style=thin edge] (33) to (36);
		\draw [style=thin edge] (28) to (34);
		\draw [style=thin edge] (28) to (35);
		\draw [style=thin edge] (28) to (36);
	\end{pgfonlayer}
\end{tikzpicture}
    \caption{
        The fishnet gadget $F_k$ for~$k=3$, connected to $S_3$ whose vertices are the three to the right, represented by dotted circles.
    }
    \label{fig:gadget-approx}
\end{figure}

A \emph{stable graph} is a graph with no edges (every vertex has degree~0). We denote by $S_k$ the stable graph on $k\geq 1$ vertices.
We define $H_k(x,y,q,d,p)$ as the graph  obtained from $G(q,d,p)$ by using the gadget depicted in \Cref{fig:gadget-approx} referred to as the \emph{fishnet} gadget~$F_k$. This gadget graph  consists of $k+2$ copies of $S_k$ connected in sequence (i.e., every vertex of the $i$th copy is connected to all vertices of the $(i-1)$th copy, and to all vertices of the $(i+1)$th copy). Moreover, the $j$th node of the first copy of $S_k$ is connected to the $j$th node of all the $k+1$ other copies. The connection of a fishnet gadget to the graph $G(q,d,p)$ is achieved via the $k$ nodes of the last copy of~$S_k$ (see \Cref{fig:gadget-approx}).

\medskip

\begin{figure}[tb]
    \centering
    \scalebox{0.75}{\input{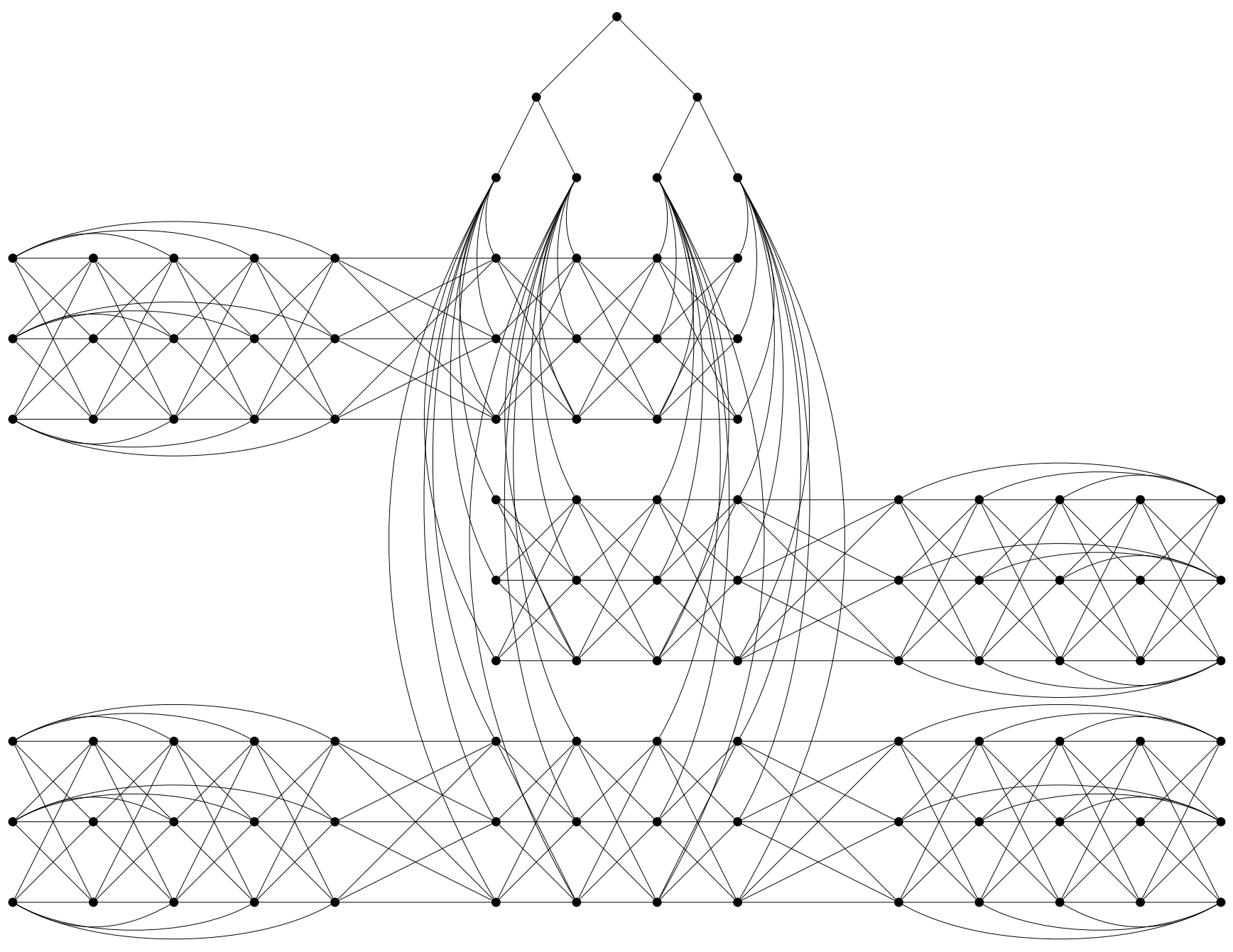}}
    \caption{
        The graph $H_k(x,y,q,d,p)$ for $k=3$, $x=(101)$, $y=(011)$,~$q=3$, $d=2$, and~$p=2$, built from~$G(3,2,2)$ shown in~\Cref{fig:gadget-3core}.
    }
    \label{fig:approx-example}
\end{figure}

\noindent The graph $H_k(x,y,q,d,p)$ is obtained from $G(q,d,p)$ as follows:
\begin{itemize}
    \item Each vertex~$u$ in a path of $G(q,d,p)$ is replaced by a copy $S^u_k$ of the stable graph~$S_k$.
    \item Each path edge~$uv$ is replaced by~$k^2$ edges connecting all vertices of~$S^u_k$ to~$S^v_k$.
    \item Each edge~$\ell u$ from a leaf~$\ell$ to a path vertex~$u$ is replaced by~$k$ edges connecting~$\ell$ to all vertices in~$S^u_k$.
    \item For each~$i \in [q]$, if~$x_i=1$ then a fishnet $F_k$ is attached to the first copy of $S_k$ in the $i$th path, and if $y_i=1$ then a fishnet is attached to the last copy of $S_k$ in the $i$th path.
\end{itemize}
\Cref{fig:approx-example} provides an example of $H_k(x,y,q,d,p)$.

\medskip

The interest of the graph $H_k(x,y,q,d,p)$ is that its degeneracy differs a lot depending on whether $x$ and $y$ are disjoint.

\begin{lemma}\label{res:hk-core}
    The degeneracy $d_k$ of $H_k(x,y,q,d,p)$ satisfies $d_k\leq k+1$ if $x$ and $y$ are disjoint, and $d_k \geq 2k$ otherwise.
\end{lemma}

\begin{proof}
    Let us first assume that $x$ and $y$ are disjoint, and let us assume for the purpose of contradiction that $H_k(x,y,q,d,p)$ contains a nonempty $(k+2)$-core~$H'$. Note that $H'$ cannot be contained entirely in a fishnet $F_k$ because $F_k$ has minimum degree $\leq k+1$ if it is not attached to the graph $G(q,d,p)$. Similarly $H'$ cannot be entirely contained in the tree $T$ of $G(q,d,p)$. As a consequence, $H'$~must contain a vertex $u$ of some copy of $S_k$ in some path. Let us say that $u$ belongs to the $j$th copy of $S_k$ in the $i$th path, for some $i \in [q]$ and $j \in [d^p]$. On the one hand, node $u$ has at least $k+2$ neighbors in~$H'$. On the other hand, in~$H_k(x,y,q,d,p)$, node $u$ has one incident edge connecting it to the tree $T$, at most $k$~neighbors to its left\footnote{In the~$(j-1)$th copy of~$S_k$ in the~$i$th path if~$j > 1$, or in the fishnet if~$j=1$ and~$x_i=1$.} and, analogously, at most~$k$ neighbors to its right. It follows that, in $H'$, $u$ must have at least one neighbor to each side. Repeating this argument yields a contradiction, since~$x_i = 0$ or~$y_i = 0$, meaning at least one of the endpoints of the~$i$th path is not attached to a fishnet. As a consequence, the $(k+2)$-core of $H_k(x,y,q,d,p)$ is empty, and thus $d_k \leq k+1$.

    Conversely, if $x$ and $y$ are not disjoint, then there exists~$i \in [q]$ such that $x_i = y_i = 1$. Let $H'$ be the subgraph of $H_k(x,y,q,d,p)$ consisting of all copies of $S_k$ in the $i$th path, and the fishnets attached to its first and last copies of~$S_k$. Note that $H'$~has minimum degree~$2k$ because, when connected to $H_k(x,y,q,d,p)$, a fishnet has minimum degree~$2k$. This implies that the $2k$-core of $H_k(x,y,q,d,p)$ is not empty, and thus $d_k\geq 2k$.
\end{proof}

\begin{lemma}\label{res:disj-to-approx-degen}
    If there exists an algorithm~$\cA$ solving \DegenApprox on $n$-node networks with diameter $\cO(p+\nicefrac{1}{\varepsilon})$ in $R(n)= \cO(n)$ rounds  under \CONGESTB, then there exists an algorithm $\cB$ solving $\Disj{q}{s}{r}$ on $G(q,d,p)$ in $\cO(\frac{1}{\varepsilon^2} R(\frac{q}{\varepsilon^2} d^p))$ rounds under \CONGESTB. Moreover, if $\cA$ is a randomized Monte-Carlo algorithm succeeding with probability $\nicefrac23$, then so is~$\cB$.
\end{lemma}

\begin{proof}
    Suppose that there is an algorithm~\cA for~\DegenApprox as in the statement. For designing~$\cB$,  let $k = \Theta(\nicefrac{1}{\varepsilon})$ to be specified later. We proceed as in the proof of~\Cref{res:disj-to-degen}, merely using $H_k(x,y,q,d,p)$ instead of $H(x,y,q,d,p)$. In particular, the information needed to determine how to simulate $H_k(x,y,q,d,p)$, which depends only on $x$ and $y$ can be sent by $r$ and $s$ to their neighbors in a single round. Furthermore, the nodes can simulate the virtual graph $H_k(x,y,q,d,p)$ on the actual network $G(q,d,p)$. Specifically, each vertex of $G(q,d,p)$ needs to simulate $\cO(k^2)$ nodes, and each edge of $G(q,d,p)$ is mapped to $\cO(k^2)$ edges of $H_k(x,y,q,d,p)$. So, each round of $\cA$ can be simulated in $\cO(k^2)$ rounds of~$\cB$. Note that $H_k(x,y,q,d,p)$ has $\cO(k^2 q + k q d^p)$ nodes (and~$\cO(k^2 q d^p)$ edges), and therefore it has $\cO(k^2 qd^p)$ nodes. Also, it has diameter $\cO(p+k)$. Therefore, the total running time of $\cB$ is $\cO(k^2 R(k^2 q d^p)) = \cO(\frac{1}{\varepsilon^2} R(\frac{q}{\varepsilon^2} d^p))$.

    An appropriate choice of $k$ is $\lfloor \frac{2-\varepsilon}{\varepsilon} \rfloor + 1$ because this choice guarantees that $\frac{2k}{k+1} > 2-\varepsilon$. Let~$c_v$ be any~$(2-\varepsilon)$-approximation to the degeneracy~$d_k$ of~$H_k(x,y,q,d,p)$ computed by any node, that is,~$d_k \leq c_v \leq (2-\varepsilon)d_k < \frac{2k}{k+1} d_k$. By~\Cref{res:hk-core}, if~$x$ and~$y$ are not disjoint then $d_k \geq 2k$, implying~$c_v \geq d_k \geq 2k$. If $x$ and $y$ are disjoint, and if $d_k \leq k+1$, then $2k \geq \frac{2k}{k+1} d_k > c_v$. Thus if~$c_v < 2k$ then~$x$ and~$y$ are disjoint, and otherwise they are not, proving the correctness of~$\cB$.
\end{proof}

\begin{theorem}
    Let $\varepsilon>0$, $B\geq 1$, and $p\geq 1$. Solving \DegenApprox on $n$-node networks with diameter $\cO(p+\nicefrac{1}{\varepsilon})$  requires $\Omega(\varepsilon^2 (\frac{n \varepsilon^2}{pB})^{\nicefrac{p}{2p+1}})$ rounds in the \CONGESTB model. This holds even for randomized Monte-Carlo algorithms succeeding with probability~$\nicefrac23$.
\end{theorem}

\begin{proof}
    Suppose that there is a (randomized) algorithm solving  \DegenApprox on $n$-node networks in $R(n) = \cO(n)$ rounds. By~\Cref{res:disj-to-approx-degen}, there is a (randomized) algorithm for~$\Disj{q}{s}{r}$ on $G(q,d,p)$ running in~$\cO(\frac{1}{\varepsilon^2} R(\frac{q}{\varepsilon^2} d^p))$ rounds. By~\Cref{res:disj-lower-bound}, we get $R(\frac{q}{\varepsilon^2} d^p) = \Omega(\varepsilon^2 d^p)$, when~$q = d^{p+1} p B$.
    Now, for the appropriate $k = \Theta(\nicefrac{1}{\varepsilon})$, the number of nodes in $H_k(x,y,q,d,p)$ is $n = \cO(k^2 q d^p) = \cO(\frac{1}{\varepsilon^2} d^{2p+1} p B)$. As $R(n) = \Omega(\varepsilon^2 d^p)$, expressing $d^p$ in terms of $n$ yields~$R(n) = \Omega(\varepsilon^2 (\frac{n \varepsilon^2}{pB})^{\nicefrac{p}{2p+1}})$, as claimed.
\end{proof}

\begin{corollary}
    For any fixed\footnote{Not depending on~$n$.} $\varepsilon>0$, solving  \DegenApprox on $n$-node networks of diameter $\cO(\log n)$ requires $\Omega(\sqrt{\nicefrac{n}{\log^2 n}})$ rounds in \CONGEST, and solving  \DegenApprox on $n$-node networks \replaced[id=cris]{with}{of} constant diameter requires $\Omega(\sqrt[3]{\nicefrac{n}{\log n}})$ rounds.
\end{corollary}

\section{Conditional Lower Bounds}
\label{sec:conditional}

Based on the lower bounds from~\Cref{sec:moving-cuts-lb}, one might be tempted  to design an~$\cO(\sqrt{n})$, or even an~$\cO(n^{0.99})$-round algorithm for degeneracy in~\CONGEST.
However, in this section, we show that, under modest restrictions, the existence of such an algorithm would imply major breakthroughs in parallel complexity theory.

Recall that a sequential algorithm $\mathcal{A}$ running in time $t(n)= \Omega(n)$ on inputs of size~$n$  admits a \emph{polynomial parallel speedup}~\cite{KruskalRS1990} if there exist $\varepsilon>0$ and a \pram\ algorithm~$\mathcal{B}$ that implements $\mathcal{A}$ so that~$\mathcal{B}$ runs in time $T(n)$  using polynomially many processors, where~${t(n)/T(n) = \Omega(n^\varepsilon)}$.
We relate the difficulty of solving the degeneracy problem in \CONGEST with the difficulty of solving its decision version in parallel, for which many hardness results are known.
In particular, the trivial sequential peeling algorithm for degeneracy takes~$\cO(n)$ time. Moreover, the problem is~$\P$-complete, that is, a~$\polylog(n)$ time parallel algorithm is unlikely to exist. In fact, no~$\cO(n^{1-\varepsilon})$ time parallel algorithm is known, for any~$\varepsilon > 0$, that is, degeneracy is not known to admit a polynomial parallel speedup. The same is the case for the aforementioned circuit value problem (\CVP). We refer the reader to~\cite{greenlaw_limits_1995} for an in-depth discussion on \P-completeness and parallelization.

\subsection{Simulation Lemma}

An $R(n)$-round \CONGEST algorithm has  local parallel time $T(n)$ if it terminates in at most $R(n)$ rounds and, at every node, and at every round, the node runs in time  $T(n)$ using $\poly(n)$ processors on a \pram for computing its new states and the messages to be sent. Let~$\pi$ be a decision problem in graphs (e.g., \kDeg[2]). For solving $\pi$ in \CONGEST, each node must output~\emph{accept} or~\emph{reject} for the problem, with global interpretation being that all nodes must accept if and only if $\pi$ holds. In the \pram model the algorithm must simply accept or reject, according to whether $\pi$ holds or not.

\begin{lemma}\label{lem:CONGESTtoPRAM}
    Let \cG be a class of graphs, and let~$\pi$ be a decision problem in graphs. If there exists a \CONGEST algorithm for~$\pi$ in $\mathcal{G}$, running on $n$-node graphs in $R(n)$ rounds, with $T(n)$ local parallel computation time in the \pram model, then there exists a \pram algorithm for~$\pi$ in $\mathcal{G}$ that runs in $\widetilde{\cO}(R(n)\cdot T(n))$ time using~$\poly(n)$ processors.
\end{lemma}

\begin{proof}
    Let $\cA$ be a \CONGEST algorithm for~$\pi$ in \cG, as in the statement. For sufficiently large~$n$, there exists a constant $c>0$ such that, when running on an $n$-node network, each node uses at most~$n^{c}$ processors.
    We split each round of~$\cA$ into two phases: computation, and communication. At the beginning of round~$t$, each node $u$ is in a memory state $s_u^t$, and holds a list of messages~$I^{t}_u = \{m_{wu}^{t}: w \in N(u)\}$ received from its neighbors. Initially, all nodes are in the same empty memory state, and all lists of messages are empty. In the computation phase of round~$t$, node~$u$ uses $s_u^t$ and $I_u^t$ for computing a set of messages~$O^{t}_u = \{m_{uw}^{t+1} : w \in N(u)\}$, to be sent to its neighbors, and updates its memory state from $s_u^t$ to~$s_u^{t+1}$. In the communication phase, the messages are delivered.

    We define the desired \pram algorithm $\cB$ as follows. Let $G\in \cG$ be an $n$-node instance of~$\pi$. Algorithm~$\cB$ stores $s^t_u$, $I^{t}_u$, and $O^{t}_u$, and uses $n^c$ processors for each node $u$ of $G$. In addition, $\cB$ uses $n^2$ other processors, for simulating the communication phase of~$\cA$.
    \begin{itemize}
        \item During the computation phase of round~$t$, $\cB$ emulates (in parallel) the local computation of the \CONGEST algorithm~$\cA$. Each node~$u$, starting from the memory state~$s_u^t$, and using the incoming messages in~$I^{t}_u$, computes~$O^{t}_u$ and $s^{t+1}_u$ in $T(n)$ time using its~$n^c$ processors. The \pram algorithm~$\cB$ uses~$n\cdot n^c$ processors for simulating this phase.

        \item During the communication phase of round $t$, processor~$P_{uw}$ overwrites~$m^t_{uw}$ from~$I^t_w$ with~$m^{t+1}_{uw}$, which takes~$\cO(\log n)$ time. Overall, this process transforms the stored~$I^t_w$ into~$I^{t+1}_w$ for all~$w$. After this,~$\cB$  is ready to start the simulation of round~$t+1$ of the \CONGEST algorithm~$\cA$.
    \end{itemize}
    At the end of the simulation (i.e., when all nodes have terminated), $\cB$ checks the output of each node, and accepts if and only if all nodes have accepted. This operation can be performed by~$\cB$ in~$\cO(\log n)$ time. In total,~$\cB$ uses $n^{c+1}+n^2 = \poly(n)$ processors, and takes $\Ot(T(n))$ time to (1)~simulate each computation phase, (2)~deliver the messages in each communication phase, and (3)~check whether all nodes in the network have terminated the execution of~$\cA$. It follows that $\pi$ in $\mathcal{G}$ admits a \pram algorithm that, on $n$-node graphs from $\mathcal{G}$, runs in $\Ot(R(n) \cdot T(n))$ time using $\poly(n)$  processors.
\end{proof}

As a consequence, the existence of a ``fast'' algorithm for~$\pi$ in \CONGEST implies the existence of a ``fast'' \pram algorithm for~$\pi$. This fact allows us to translate known hardness results about degeneracy from the parallel computing context to \CONGEST. This is what we do next.

\subsection{Conditional Lower Bound for \texorpdfstring{\kDeg[2]}{2-Degenerate}}
\label{sec:condbds}

Let~$\pi_1$ and~$\pi_2$ be two graph decision problems. We say~$\pi_1$ has a \emph{quasi-linear \NC-reduction} to~$\pi_2$, denoted by $\pi_1 \leq^{\NC}_{\mbox{\tiny quasi}} \pi_2$, if there is a function~$f$ from the set of graphs to the set of graphs such that, for every graph~$G$,
\begin{itemize}
    \item $\pi_1$ accepts~$G$ if and only if $\pi_2$ accepts~$f(G)$,
    \item $f$ is computable in \NC,
    \item $|E(f(G))| = \widetilde{\cO}(|E(G)|)$, and
    \item $D(f(G)) = \widetilde{\cO}(D(G))$, where~$D(H)$ denotes the diameter of graph~$H$.
\end{itemize}

\begin{lemma}\label{res:pi-lower-cvp}
    Let~$\pi$ be a decision problem such that~$\CVP \leq^{\NC}_{\mbox{\tiny quasi}}  \pi$.
    Unless \CVP admits a polynomial parallel speedup, it holds that, for every $\varepsilon \in (0,1)$ and $\delta \in (0,\varepsilon)$, $\pi$ does not admit a \CONGEST algorithm running in $\cO(n^{1- \varepsilon})$ rounds with $\cO(n^{\delta})$ local parallel computation time in $n$-node networks of diameter $\widetilde{\cO}(1)$.
\end{lemma}

\begin{proof}
    Let us assume that there exists a \CONGEST algorithm for~$\pi$ as in the statement.
    Then, by \Cref{lem:CONGESTtoPRAM}, there exists a \pram algorithm \cA for~$\pi$ that runs in $\widetilde{\cO}(n^{1-\varepsilon} \cdot n^{\delta}) = \widetilde{\cO}(n^{1 - \varepsilon + \delta})$ time using $\poly(n)$ processors.
    Consider any instance~$C$ of size~$n$ for \CVP, and modify it as follows. First, remove all pairs of consecutive NOT gates. Second, add a 0-gate connected to all OR-gates, and add a 1-gate connected to all AND-gates. The resulting circuit has size~$\cO(n)$, and constant diameter. Moreover, it has the same circuit value.
    Using the reduction from \CVP to~$\pi$, one can in~$\polylog(n)$ time construct an $\widetilde{\cO}(n)$-node graph~$G_C$ of diameter $\polylog(n)$ that is an instance of~$\pi$ such that $\pi$ accepts $G_C$ if and only if the circuit value of $C$ is~1.
    One can solve the instance~$G_C$ with the \pram algorithm~$\cA$, which runs in~$\widetilde{\cO}(n^{1-\varepsilon+\delta})$ time using $\poly(n)$ processors. This implies that \CVP admits a polynomial parallel speedup, as ${1 - \varepsilon + \delta < 1}$.
\end{proof}

\begin{theorem}\label{cor:3core-lower-cvp}
    Let $\varepsilon \in (0,1)$ and $\delta \in (0,\varepsilon)$. Unless \CVP admits a polynomial parallel speedup, \kDeg[2] does not admit a \CONGEST algorithm running in $\cO(n^{1- \varepsilon})$ rounds with $\cO(n^{\delta})$ local parallel computation time on $n$-node networks, even on graphs of maximum degree~4 and diameter $\cO(\log n)$.
\end{theorem}

\begin{proof}
    Consider the monotone fan-in~2 fan-out~2 restriction to the circuit value problem, referred to as \MCVP.
    From any instance $C$ of \CVP, one can construct an instance $C'$ of \MCVP by applying the standard reduction that removes all NOT gates, and uses balanced binary trees to decrease the maximum out- and in-degree to~2 (see~\cite[Theorem 6.2.3]{greenlaw_limits_1995}). If~$C$ has size~$n$ and diameter~$D$, then~$C'$ has size~$\cO(n)$ and diameter~$\cO(D \log n)$. Then, we use the reduction from~\MCVP to~\kDeg[2] from~\cite[Theorem 3]{anderson_p-complete_1984}, which we refrain to present due to space restrictions. It produces a graph~$G$ with~$\cO(n)$ nodes, maximum degree 4, and diameter~$\cO(D \log n)$. It follows that $\CVP \leq^{\NC}_{\mbox{\tiny quasi}} \kDeg[2]$, and the result follows thanks to \Cref{res:pi-lower-cvp}.
\end{proof}

\subsection{Remarks on the Local Computation Time}
\label{sec:remarks}

The conditional lower bounds on the round complexity of~\kDeg[2] in \CONGEST derived in \Cref{sec:condbds} use \Cref{res:pi-lower-cvp} with local computation time $\cO(n^{\delta})$ for any $\delta < \varepsilon$. Thus it allows the existence of an algorithm running in~$\cO(n^{1-\varepsilon})$ rounds if it has~$\Omega(n^\varepsilon)$ local computation time. Nevertheless, it is worth noticing that such local computation time is significantly larger than the one of a typical \CONGEST algorithm.
This subsection argues that an $\cO(n^{\delta})$ upper bound on the local computation may in fact be considered quite light,
as far as the \CONGEST model is concerned.

\begin{itemize}
    \item First, every node receives $\cO(\log n)$ bits on each of its incident edges per round. Thus, in networks of constant degree, the total number of bits a node receives during $r$ rounds is $\cO(r\log n)$. Roughly speaking, as a function of the input size, $\cO(n^{\delta})$ time corresponds to exponential time for $r=\polylog(n)$, and to sublinear time only for $r = \omega(n^{\delta}/\log n)$.

    \item Second, in networks with constant degree, for most known algorithms, the local memory state of each node can be encoded in $\polylog(n)$ bits throughout their whole execution, and the computation performed at each node is often elementary (e.g., computing the minimum or maximum of $\cO(\log n)$ integers, performing a constant number of XOR operations on $\cO(\log n)$-bit strings, etc.). This holds even for non-trivial problems such as leader election, maximal independent set, maximal matching, and minimum-weight spanning tree (see, e.g., \cite{KuttenP98,Linial92,Peleg2000}).

    \item Last but not least, in our simulation lemma, the computation time $T(n)$ used by each node per round may not only be measured assuming the sequential \ram model, but may assume the stronger parallel \ram (\pram) model. That is, we allow each node to perform an amount of computation that takes $T(n)$ time on a \pram with $\poly(n)$ processors. Even for~${T(n) = \polylog(n)}$, each node can perform a significant amount of computation, such as prefix sums and matrix operations.
\end{itemize}

\subsection{Lower Bound for other P-Complete Problems}
\label{subsubsec:other-P-complete}

Many~\P-complete problems~$\pi$ satisfy~$\CVP \leq^{\NC}_{\mbox{\tiny quasi}} \pi$. The same lower bound as in \Cref{cor:3core-lower-cvp} can be derived for these problems.
This is for instance the case of \DegenApprox, and of other variants of~\kCore such as directed $k$-cores, and $k$-truss, which is the variant of $k$-core in which each edge in the $k$-truss must be part of $k-2$ triangles. This follows from the known quasi-linear reductions from monotone circuit value to all of these problems~\cite[\S 3]{couto_hardness_2025}.
In fact, several reductions presented in~\cite{greenlaw_limits_1995} start from \CVP, and are quasi-linear. \Cref{res:pi-lower-cvp} thus yields other results such as the following, with the same proof as for \Cref{cor:3core-lower-cvp}.

\begin{theorem}\label{res:more_conditional}
    Let $\varepsilon \in (0,1)$ and $\delta \in (0,\varepsilon)$. Unless \CVP admits a polynomial parallel speedup, none of the problems \LFMIS, \LFDVC, and \MaxFlow
    admit a \CONGEST algorithm  running in $\cO(n^{1- \varepsilon})$ rounds with $\cO(n^{\delta})$ local parallel computation time in $n$-node networks of diameter $\widetilde{\cO}(1)$.
\end{theorem}

\MaxFlow has a known unconditional lower bound of~${\widetilde{\Omega}(D + \sqrt{n})}$~\cite[\S 4]{sarma_distributed_2011}. Our  lower bound is much higher, but is conditional.
\LFMIS and \LFDVC, which are classic problems in parallel complexity literature~\cite{greenlaw_limits_1995}, had no previous lower bounds in \CONGEST.

\begin{remark*}
    We note that our unconditional lower bound $\widetilde{\Omega}(\sqrt{n})$ for \Degen (\Cref{cor:kdeg2-lb}) can be extended to other problems, such as the circuit value problem or \MCVP. (Due to space restrictions, we do not explain these particular extensions.)
\end{remark*}

The remark above has consequences. Indeed, the circuit value is used as a base problem to prove the \P-completeness of many other problems, so our lower bound can also be extended to the \CONGEST version of many~\P-complete problems. Specifically, the unconditional lower bounds can be extended to a (\P-complete) graph problem~$\pi$ if there exists a reduction from \MCVP to~$\pi$ which roughly preserves the graph structure, so that the ``simulation argument'' of~\Cref{res:disj-to-degen} still works. This is the case for several variants of~$k$-core, using the reductions from~\cite{couto_hardness_2025}, as well as for~\LFMIS and~\LFDVC, using the standard reductions from~\cite{greenlaw_limits_1995}. This is because all these problems have the following property: in their reduction from \MCVP, each gate~$g$ is mapped to a (constant-size) subgraph~$H_g$, and if~$H_g$ has an edge to~$H_f$ in the final graph, then~$g$ is connected to~$f$ in the original circuit.

\section{The Challenge of Finding Better Unconditional Lower Bounds}
\label{sec:upper-bounds}

In this section, we discuss the ability to derive better (unconditional) lower bounds for the problems considered in this paper, above the apparent barrier of $\widetilde{\Omega}(\sqrt{n})$. We show that recent tools for deriving lower bounds in \CONGEST will not suffice.

\subsection{On the Limitation of Moving Cuts}

Das Sarma et al.~\cite{sarma_distributed_2011} were the first to analyze the complexity of distributed algorithms on the graphs $G(q, d, p)$, and derive lower bounds for \CONGEST from lower bounds for communication complexity problems such as disjointness. In more recent literature~\cite{haeupler_network_2020}, this technique has been dubbed ``moving cuts'' by Haeupler, Wajc, and Zuzic~\cite{haeupler_universally-optimal_2021} who have generalized this technique, examining exactly which properties of the graph~$G(q, d, p)$ make disjointness hard to solve.
Given any graph $G$, the parameter $\MovingCut(G)$ is defined as the maximum distance $\beta$ for which there exists a set $S = \{(s_i, t_i), i\in\{1,\dots,k\}\}$ of connectable source-sink pairs  (i.e., vertex pairs that can be connected by vertex-disjoint paths in~$G$) and an edge-length assignment $\ell: E(G) \to \mathbb{N}_{\ge 1}$ satisfying:
\[
    \sum_{e \in E} (\ell(e) - 1) < k, \;\; \mbox{and, for every $i,j \in \{1,\dots, k\}$,}\;\;
    \text{dist}_\ell(s_i, t_j) \geq \beta.
\]
For the~$n$-vertex graph $G = G(\cO(\sqrt{n}), 2, \cO(\log n))$, we have
\[
    \MovingCut(G) = \widetilde{\Omega}(\sqrt{n})\text{~\cite{haeupler_universally-optimal_2021}}.
\]
This lower bound is the source of the difficulty of computing~$\Disj{\beta}{s}{r}$ in $G$ (cf.\ \Cref{res:disj-lower-bound}). For several \CONGEST problems, the framework in~\cite{haeupler_universally-optimal_2021} enables one to derive the lower bound of $\widetilde{\Omega}(\MovingCut(G))$ rounds for every graph~$G$. Nevertheless, this lower bound does not hold for~\Degen. Indeed, the distributed peeling algorithm for \kDeg[2] takes~$\cO(p)$ rounds in the graph $G(q, d, p)$\footnote{In the first~$p$ rounds, all nodes from the tree are peeled top to bottom; After that, all nodes in the $q$ paths have degree at most two, and are thus peeled on the next round.}, which has diameter~$2p+2$. As a consequence, $\widetilde{\Omega}(\MovingCut(G))$ is not a lower bound for the complexity of~\Degen in \CONGEST.

\begin{remark*}
    Haeupler, Wajc, and Zuzic~\cite{haeupler_universally-optimal_2021} also designed algorithms for several problems running in $\widetilde{\cO}(\MovingCut(G))$. One might wonder whether such an algorithm exists for~\Degen. Although our unconditional lower bound does not show otherwise, \Cref{sec:conditional} provides strong evidence that such an algorithm is unlikely to exist.
\end{remark*}

\subsection{On the Limitation of Small Cuts}

Here we prove that the framework by Abboud et al.~\cite{AbboudCKP21}, which is used to obtain many polynomial lower bounds in \CONGEST, cannot be directly used to obtain any non-trivial lower bound for~\Degen. We start with a review of the framework from~\cite{AbboudCKP21}.

\subsubsection{The Framework}\label{sec:cc-defs}

A \emph{two-party communication problem}~\cite{rao_communication_2020} is specified by a function $f : A \times B \to \{0,1\}$. It involves two players, referred to as Alice and Bob. Alice receives an input~$a \in A$ and Bob receives input~$b \in B$. The goal of the players is to output~$f(a,b)$. The communication complexity of~$f$ is the number of bits the two players must exchange in the worst instance. It is denoted by $\CC(f)$, or $\CC^R(f)$ if randomization is allowed. The following definition is borrowed from~\cite{AbboudCKP21}.

\begin{definition}\label{def:family}
    Let~$f: A \x B \to \zo$ be a function, and let $P$ be a graph predicate. A family of graphs~$\left\{G_{x,y} = (V, E_{x,y}) : (x, y) \in A \x B \right\}$
    with a partition~$V = V_A \cup V_B$, and possibly labels on the vertices and edges, is said to be a \emph{family of lower bound graphs w.r.t.~$f$ and~$P$} if the following properties hold:
    \begin{enumerate}
        \item only the edges in~$G[V_A]$ and their labels may depend on~$x$;
        \item only the edges in $G[V_B]$ and their labels may depend on~$y$;
        \item $f(x,y)$ is uniquely determined by whether $G_{x, y}$ satisfies the predicate~$P$.
    \end{enumerate}
\end{definition}

Abboud et al.~\cite{AbboudCKP21} proved the following theorem, and derived (unconditional) lower bounds in \CONGEST from it, for several problems such as exact diameter and vertex cover.

\begin{theorem}[Theorem 6 in~\cite{AbboudCKP21}] \label{thm:lower-no-round}
    Let $f: \zo^K \x \zo^K \to \zo$ be a function, and let $P$ be a graph predicate. If there exists a family~$\{G_{x,y} = (V, E_{x,y}): (x, y) \in \zo^K \x \zo^K \}$ of lower bound graphs w.r.t.~$f$ and~$P$ with ${V = V_A \cup V_B}$ and~$C = E[V_A, V_B]$, then every distributed algorithm for deciding~$P$  requires~$\Omega(\CC(f)/|C|\cdot \log |V|)$ rounds in the \CONGEST model, and every randomized algorithm requires~$\Omega(\CC^R(f)/|C|\cdot \log |V|)$ rounds.
\end{theorem}

\subsubsection{Why the Framework of Small Cuts Fails for \Degen}

Let $f : A \times B \rightarrow \{0,1\}$ be a function. If it is possible to use an algorithm for~\kDeg to decide, given any $(x,y) \in A \times B$, whether $f(x,y) = 1$, that is, if there exists  a family of lower bound graphs w.r.t.\ $f$ and \kDeg, then we will show an efficient communication protocol for~$f$. This implies that the communication complexity~$\CC(f)$ is low and, hence, \Cref{thm:lower-no-round} yields only trivial lower bounds for~\kDeg. Recall that, in the peeling algorithm, the vertices with degree at most~$k$ are progressively removed from the input graph~$G$, until no vertices have degree at most~$k$. In the end, the remaining graph is precisely the $(k+1)$-core of~$G$, and~$G$ is $k$-degenerate if and only if the $(k+1)$-core is empty. The order in which the vertices are peeled (which may not be unique) is called a \emph{peeling order for the $(k+1)$-core}. With this in mind, we can show the communication protocol.

\begin{figure}[tb]
    \centering
    \begin{tikzpicture}
	\begin{pgfonlayer}{nodelayer}
        \node [] at (-1,0) {(a)};
		\node [style=dashed circle] (0) at (0, 0) {};
		\node [style=onegate] (1) at (1, 0) {};
		\node [style=onegate] (2) at (2, 0.5) {};
		\node [style=onegate] (3) at (2, -0.5) {};
		\node [style=onegate] (4) at (2, 0) {};
		\node [style=onegate] (5) at (3, 0) {};
		\node [above=1pt of 1] (6) {$u$};
		\node [above=1pt of 0] (7) {$a$};
	\end{pgfonlayer}
	\begin{pgfonlayer}{edgelayer}
		\draw (1) to (2);
		\draw (1) to (3);
		\draw (3) to (4);
		\draw (4) to (2);
		\draw (2) to (5);
		\draw (5) to (3);
		\draw (4) to (5);
		\draw [style=dashed edge] (0) to (1);

        \node [] at (5,0) {(b)};
		\node [style=dashed circle] (0) at (6, 0) {};
		\node [style=onegate] (1) at (7, 0) {};
		\node [style=onegate] (2) at (8, 0.7) {};
		\node [style=onegate] (3) at (8, -0.7) {};
		\node [style=onegate] (4) at (8, 0) {};
		\node [style=onegate] (5) at (9, 0.5) {};
       	\node [style=onegate] (6) at (9, -0.5) {};
		\node [above=1pt of 1] {$u$};
		\node [above=1pt of 0] {$a$};
	\end{pgfonlayer}
	\begin{pgfonlayer}{edgelayer}
		\draw (1) to (2);
		\draw (1) to (3);
		\draw (3) to (4);
		\draw (4) to (2);
		\draw (2) to (5);
		\draw (5) to (3);
		\draw (4) to (5);
		\draw (2) to (6);
		\draw (6) to (3);
		\draw (4) to (6);
		\draw (5) to (6);
		\draw (1) to (4);
		\draw [style=dashed edge] (0) to (1);

	\end{pgfonlayer}
\end{tikzpicture}
    \caption{(a) Gadget used in \Cref{thm:upper-bound-communication} for~\kDeg[2]. In any peeling order for the~$3$-core, vertex~$a$ is always the first to be removed (if it is removed), independently from the rest of the graph. Furthermore, if~$a$ is removed, then the rest of the gadget will be removed as well. (b) Similar gadget for~\kDeg[3].}
    \label{fig:arrow}
\end{figure}
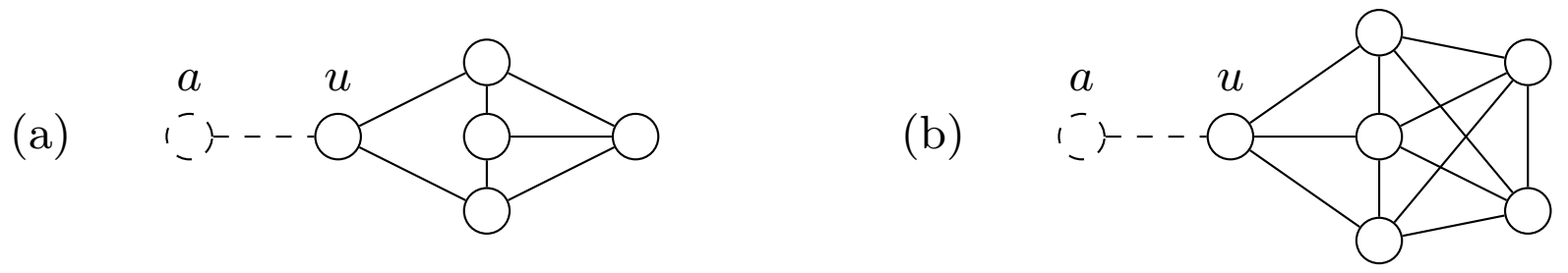

\newcommand{\Gr}[1]{\ensuremath{\widetilde{G}_{#1}}}

\begin{theorem}\label{thm:upper-bound-communication}
    Let $k \geq 2$ be an integer, and let $f : A \x B \rightarrow \{0,1\}$ be a function. If there exists a family of lower bound graphs $\{G_{x,y} = (V, E_{x,y}):(x,y)\in A\times B\}$ w.r.t.\ $f$ and~\kDeg where $V = V_A \cup V_B$ and $C = E[V_A, V_B]$, then~$\CC(f) = \cO(|C| \cdot \log |V|)$.
\end{theorem}

\begin{proof}
    Let us assume that there exists  a family of lower bound graphs $\{G_{x,y} = (V, E_{x,y}):(x,y)\in A\times B\}$ as in the statement of the theorem.
    We first prove that there is a communication protocol to compute~$f(x,y)$ using~$\cO(|C| \cdot \log |V|)$ bits of communication for $k=2$, and we show later how to generalize this fact to larger~$k$.
    For any~$(x, y) \in A \x B$, let us consider $G_{x, y}$.  Let $C_A = V(C) \cap V_A$ and $C_B = V(C) \cap V_B$. Moreover, let \Gr{A} be the graph $G[V_A]$ and, for every $ab \in C$ such that $a \in C_A$ and $b \in C_B$, add a copy $L_{ab}$ of the graph in Figure~\ref{fig:arrow}(a) to~\Gr{A}, connecting~$a \in C_A$ to~$u \in V(L_{ab})$. Define~\Gr{B} similarly. See~\Cref{fig:cc_upper} for an example.
    We now describe a communication protocol between Alice and Bob to determine whether~$G$ is~$2$-degenerate, and thus to compute~$f(x,y)$, by simulating the peeling algorithm. This protocol is described as Algorithm~\ref{alg:protocol}.

    \begin{figure}[tb]
        \centering
        \captionsetup[subfigure]{justification=centering}
        \begin{subfigure}{0.49\textwidth}
            \centering
            \begin{tikzpicture}
	\begin{pgfonlayer}{nodelayer}
		\node [style=none] (0) at (-8, 4.5) {};
		\node [style=none] (1) at (-8, 0.5) {};
		\node [style=none] (2) at (-6, 0.5) {};
		\node [style=none] (3) at (-6, 4.5) {};
		\node [style=none] (4) at (-4, 4.5) {};
		\node [style=none] (5) at (-4, 0.5) {};
		\node [style=none] (6) at (-2, 0.5) {};
		\node [style=none] (7) at (-2, 4.5) {};
		\node [style=onegate] (8) at (-6.25, 4) {};
		\node [style=onegate] (9) at (-6.25, 3) {};
		\node [style=onegate] (10) at (-6.25, 2) {};
		\node [style=onegate] (11) at (-6.25, 1) {};
		\node [style=onegate] (12) at (-3.75, 4) {};
		\node [style=onegate] (13) at (-3.75, 2.5) {};
		\node [style=onegate] (14) at (-3.75, 1) {};
		\node [style=none] (15) at (-6.75, 4) {$a$};
		\node [style=none] (16) at (-6.75, 3) {$b$};
		\node [style=none] (17) at (-6.75, 2) {$c$};
		\node [style=none] (18) at (-6.75, 1) {$d$};
		\node [style=none] (19) at (-3.25, 4) {$e$};
		\node [style=none] (20) at (-3.25, 2.5) {$f$};
		\node [style=none] (21) at (-3.25, 1) {$g$};
		\node [style=none] (22) at (-7, 4.75) {$G[V_A]$};
		\node [style=none] (23) at (-3, 4.75) {$G[V_B]$};
		\node [style=none] (24) at (-5, 4.75) {$E[V_A, V_B]$};
	\end{pgfonlayer}
	\begin{pgfonlayer}{edgelayer}
		\node [style=graph region, fit = (3) (0) (1) (2)] {};
		\node [style=graph region, fit = (7) (4) (5) (6)] {};
		\draw (8) to (12);
		\draw (8) to (13);
		\draw (9) to (12);
		\draw (10) to (13);
		\draw (11) to (13);
		\draw (11) to (14);
	\end{pgfonlayer}
\end{tikzpicture}
            \captionsetup{textformat=simple}
            \subcaption{Graph~$G$}
        \end{subfigure}%
        \begin{subfigure}{0.49\textwidth}
            \centering
            \begin{tikzpicture}
	\begin{pgfonlayer}{nodelayer}
		\node [style=none] (0) at (-2, 3.75) {};
		\node [style=none] (1) at (-2, 0.5) {};
		\node [style=none] (2) at (0, 0.5) {};
		\node [style=none] (3) at (0, 3.75) {};
		\node [style=onegate] (4) at (-0.25, 3.25) {};
		\node [style=onegate] (5) at (-0.25, 2.5) {};
		\node [style=onegate] (6) at (-0.25, 1.75) {};
		\node [style=onegate] (7) at (-0.25, 1) {};
		\node [style=none] (8) at (-0.75, 3.25) {$a$};
		\node [style=none] (9) at (-0.75, 2.5) {$b$};
		\node [style=none] (10) at (-0.75, 1.75) {$c$};
		\node [style=none] (11) at (-0.75, 1) {$d$};
		\node [style=none] (12) at (-1, 4) {$G[V_A]$};
		\node [style=hollow] (13) at (0.75, 4) {};
		\node [style=hollow] (14) at (1.5, 4.25) {};
		\node [style=hollow] (15) at (1.5, 3.75) {};
		\node [style=hollow] (16) at (1.5, 4) {};
		\node [style=hollow] (17) at (2.25, 4) {};
		\node [style=hollow] (18) at (0.75, 3.25) {};
		\node [style=hollow] (19) at (1.5, 3.5) {};
		\node [style=hollow] (20) at (1.5, 3) {};
		\node [style=hollow] (21) at (1.5, 3.25) {};
		\node [style=hollow] (22) at (2.25, 3.25) {};
		\node [style=hollow] (23) at (0.75, 2.5) {};
		\node [style=hollow] (24) at (1.5, 2.75) {};
		\node [style=hollow] (25) at (1.5, 2.25) {};
		\node [style=hollow] (26) at (1.5, 2.5) {};
		\node [style=hollow] (27) at (2.25, 2.5) {};
		\node [style=hollow] (28) at (0.75, 1.75) {};
		\node [style=hollow] (29) at (1.5, 2) {};
		\node [style=hollow] (30) at (1.5, 1.5) {};
		\node [style=hollow] (31) at (1.5, 1.75) {};
		\node [style=hollow] (32) at (2.25, 1.75) {};
		\node [style=hollow] (33) at (0.75, 1) {};
		\node [style=hollow] (34) at (1.5, 1.25) {};
		\node [style=hollow] (35) at (1.5, 0.75) {};
		\node [style=hollow] (36) at (1.5, 1) {};
		\node [style=hollow] (37) at (2.25, 1) {};
		\node [style=hollow] (38) at (0.75, 0.25) {};
		\node [style=hollow] (39) at (1.5, 0.5) {};
		\node [style=hollow] (40) at (1.5, 0) {};
		\node [style=hollow] (41) at (1.5, 0.25) {};
		\node [style=hollow] (42) at (2.25, 0.25) {};
		\node [style=none] (43) at (2.75, 4) {$L_{ae}$};
		\node [style=none] (44) at (2.75, 3.25) {$L_{af}$};
		\node [style=none] (45) at (2.75, 2.5) {$L_{be}$};
		\node [style=none] (46) at (2.75, 1.75) {$L_{cf}$};
		\node [style=none] (47) at (2.75, 1) {$L_{df}$};
		\node [style=none] (48) at (2.75, 0.25) {$L_{dg}$};
	\end{pgfonlayer}
	\begin{pgfonlayer}{edgelayer}
		\node [style=graph region, fit=(3) (0) (1) (2)] {};
		\draw (13) to (15);
		\draw (15) to (17);
		\draw (16) to (15);
		\draw (14) to (16);
		\draw (14) to (17);
		\draw (13) to (14);
		\draw (4) to (13);
		\draw (18) to (20);
		\draw (20) to (22);
		\draw (21) to (20);
		\draw (19) to (21);
		\draw (19) to (22);
		\draw (18) to (19);
		\draw (23) to (25);
		\draw (25) to (27);
		\draw (26) to (25);
		\draw (24) to (26);
		\draw (24) to (27);
		\draw (23) to (24);
		\draw (28) to (30);
		\draw (30) to (32);
		\draw (31) to (30);
		\draw (29) to (31);
		\draw (29) to (32);
		\draw (28) to (29);
		\draw (33) to (35);
		\draw (35) to (37);
		\draw (36) to (35);
		\draw (34) to (36);
		\draw (34) to (37);
		\draw (33) to (34);
		\draw (38) to (40);
		\draw (40) to (42);
		\draw (41) to (40);
		\draw (39) to (41);
		\draw (39) to (42);
		\draw (38) to (39);
		\draw (4) to (18);
		\draw (5) to (23);
		\draw (6) to (28);
		\draw (7) to (33);
		\draw (7) to (38);
		\draw (16) to (17);
		\draw (21) to (22);
		\draw (26) to (27);
		\draw (31) to (32);
		\draw (36) to (37);
		\draw (41) to (42);
	\end{pgfonlayer}
\end{tikzpicture}
            \captionsetup{textformat=simple}
            \subcaption{Graph~\Gr{A}}
        \end{subfigure}%
        \caption{Example of how Alice builds graph~\Gr{A} from the original graph~$G$ and her set of vertices.}
        \label{fig:cc_upper}
    \end{figure}
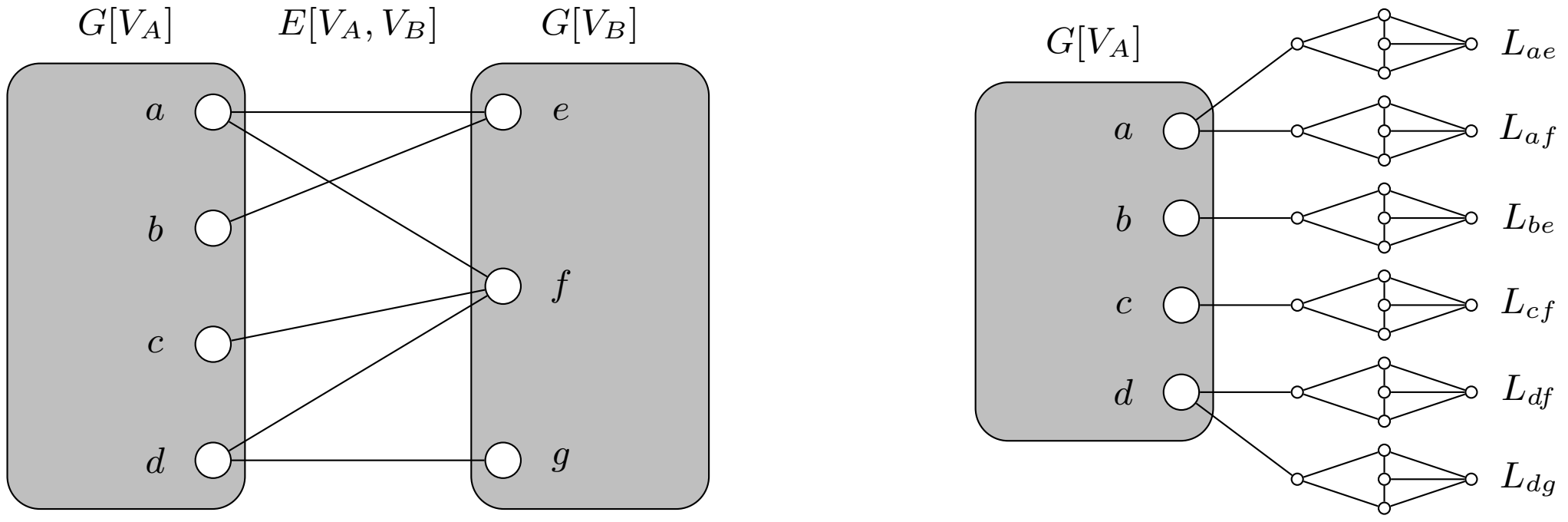

    \floatname{algorithm}{Protocol}
    \algrenewcommand\algorithmicthen{\textbf{then}}

    \begin{algorithm}[tb]
        \caption{Communication protocol for computing~$f(x,y)$.}\label{alg:protocol}
        \begin{algorithmic}[1]
            \State Alice and Bob build graphs~\Gr{A} and \Gr{B}, respectively.
            \Repeat
            \State Alice peels vertices of degree at most~$2$ from~\Gr{A} until none exists. \label{algl:Apeel}
            \State Alice sends to Bob the list of identifier pairs~$(a,b)$ for each gadget~$L_{ab}$ she peeled.
            \State For each pair~$(a, b)$ that Bob received, he removes the gadget~$L_{ba}$ from~\Gr{B}. \label{algl:Lab}
            \State Bob peels vertices of degree at most~$2$ from~\Gr{B} until none exists. \label{algl:Bpeel}
            \State Bob sends to Alice the list of identifier pairs~$(b, a)$ for each gadget~$L_{ba}$ he peeled.
            \If{Alice received a non-empty list}
            \State For each pair~$(b, a)$ that Alice received, she removes the gadget~$L_{ab}$ from~\Gr{A}. \label{algl:Lba}
            \EndIf
            \Until{Alice receives an empty list}
            \State Both players determine~$f(x,y)$
            based on whether their remaining graphs are empty.
        \end{algorithmic}
    \end{algorithm}

    \paragraph*{Total communication.}

    Each player sends~$\cO(\log |V|)$ bits per gadget peeled, and each gadget is peeled at most once, since edges can only be removed. Additionally, apart from the very first message, whenever an empty list is sent, the protocol sends at most one more empty list before terminating, using~$\cO(1)$ bits. Thus, the total number of bits exchanged between the two players is~$\cO(|C| \cdot \log |V|)$.

    \paragraph*{Correctness.}

    Let~$K_A$ and~$K_B$ be the remaining graphs for Alice and Bob, respectively, that is, their computed~$3$-cores.
    First note that~$L_{ab}$ is in~$K_A$ if and only if~$L_{ba}$ is in~$K_B$. This is because when one is peeled by lines~\ref{algl:Apeel} or~\ref{algl:Bpeel}, the other is removed in lines~\ref{algl:Lab} or~\ref{algl:Lba}, respectively. Consider the graph~$K$ formed by merging~$K_A$ and~$K_B$ and replacing gadget pairs~$L_{ab}$ and~$L_{ba}$ by the edge~$ab$. We will show that~$K$ is precisely~$K_G$, the~$3$-core of~$G$.

    Note that the degree of all vertices from~$V_A$ and~$V_B$ is the same in~$K$ as in~$K_A$ and~$K_B$, respectively, since each gadget~$L_{ab}$ contributes one to the degree of~$a$, and it was replaced by the edge~$ab$. So~$K$ has minimum degree at least~$3$ and thus~$K \subseteq K_G$. For the purpose of contradiction, let us suppose that $K_G \not\subseteq K$, and let $v$ be the first vertex in $K_G$ peeled by Alice or Bob. Let $\Gr{A}^t$ and $\Gr{B}^t$ be the graphs $\Gr{A}$ and $\Gr{B}$ at the time $t$ when $v$ was chosen to be peeled. Suppose, w.l.o.g., that $v \in V(\Gr{A}^t)$, and let us show that $d_{\Gr{A}^t}(v) \geq d_{K_G}(v)$.
    For any edge $vw \in E(K_G)$, since $v$ was the first vertex in $K_G$ to be peeled, we know $w$ remains in the graph, that is, $w \in V(\Gr{A}^t) \cup V(\Gr{B}^t)$. If $w \in V(\Gr{A}^t)$, then $vw \in E(\Gr{A}^t)$. If $w \in V(\Gr{B}^t)$, then let us show that $L_{vw} \subseteq \Gr{A}^t$. First, $L_{vw}$ was not peeled in line~\ref{algl:Apeel}, since such peeling would have started by~$v$; second, $L_{vw}$ was not removed in line~\ref{algl:Lba}, since that would imply that $w$ is peeled by Bob in order to peel $L_{wv}$ in line~\ref{algl:Bpeel}. Thus, at least 3 edges are incident to $v$ in $\Gr{A}^t$, a \highlight[id=mk,comment={I see there is a contradiction with $v$ being chosen to be peeled, but not with the fact that $K_G\not\subseteq K$}]{contradiction}. We conclude that $K_G \subseteq K$.
    This shows that the $3$-core is computed correctly by Alice and Bob, and thus whether $G$ is $2$-degenerate. Since $\left\{G_{x,y}\right\}$ is a family of lower bound graphs w.r.t.~$f$ and~\kDeg[2], this implies that $f(x, y)$ is computed correctly.

    To generalize this result to $k>2$, it suffices to change the graph from~\Cref{fig:arrow}(a) to any graph with no~$(k+1)$-core but such that all vertices have degree at least~$k+1$, except for~$v$ which has degree exactly~$k$. See~\Cref{fig:arrow}(b).
\end{proof}

It follows from \Cref{thm:lower-no-round} that, for any~$k \geq 2$, using the framework in~\cite{AbboudCKP21}, one can only obtain trivial lower bounds of~$\Omega(\CC(f)/(|C| \cdot \log |V|)) = \Omega(1)$ rounds for~\kDeg.

\begin{remark*}
    It is possible to extend the framework from~\cite{AbboudCKP21} to obtain polynomial lower bounds for~\Degen. For that, we however need to switch to the \emph{pointer chasing} problem, which has more complex lower bounds that depend on the number of alternations between the two players during the communication protocol, and which requires to modify~\Cref{thm:lower-no-round} accordingly. We can do the appropriate modifications, but this only yields a lower bound $\widetilde{\Omega}(\sqrt[3]{n})$ for~\Degen, which is weaker than the one obtained in~\Cref{sec:moving-cuts-lb}. The techniques and ideas used in the proof are however interesting, and they may be adapted to get better lower bounds for other problems. The details are given next.
\end{remark*}

\subsection{Extension of the Small Cuts Techniques}
\label{sec:cc-lower-bounds}

We established above that the framework by Abboud et al.~\cite{AbboudCKP21} cannot be directly used to obtain polynomial lower bounds for~\Degen. In this section, we extend the framework, consider the pointer chasing problem, combined with the bit gadget technique from~\cite{assadi_polynomial_2019}, in order to obtain $\widetilde{\Omega}(\sqrt[3]{n})$ lower bounds. For simplicity, we provide lower bounds for the problem~\kCore, in which each node outputs whether it is in the~$k$-core of the network.

\subsubsection{Extending the Framework}

We start by formalizing the definitions of communication complexity from~\Cref{sec:cc-defs}, in order to introduce round restrictions. A \emph{communication protocol}~$\pi$ for function $f$ is a procedure in which Alice and Bob exchange messages, where each message may depend on their input, the messages received so far, and possibly on some (public) random bits, in the randomized case. At the end, both players output $f(a, b)$.

A protocol~$\pi$ is said to run in $r$ \emph{rounds} if the players exchange messages in $r$ alternating turns, with Alice starting.  The deterministic communication complexity of~$f$ over~$r$ rounds, denoted~$\CC_r(f)$, is the minimum of~$\CC(\pi)$ over all deterministic protocols~$\pi$ for~$f$ that run in~$r$ rounds, and we define~$\CC^R_r(f)$ similarly for the randomized case.

\begin{definition}[Pointer Chasing]
    The $(\ell, m)$ history pointer chasing problem, or~\HistPC, is a two-party communication complexity problem. There are two functions ${f_A, f_B: [m] \to [m]}$ known by Alice and Bob, respectively. The goal is, for all~$i \in [\ell]$, to compute $o_i = f_i(f_{i-1}(\cdots f_1(1) \cdots))$, where $f_i$ is~$f_A$ if~$i$ is odd, and~$f_B$ otherwise.
\end{definition}

It is easy to solve~$\HistPC$ by exchanging~$\Theta(\ell \cdot \log m)$ bits over~$\ell$ rounds. Alice starts by sending~$f_A(1)$ to Bob, who then sends~$f_B(f_A(1))$ to Alice and so on. However, solving this problem with fewer rounds substantially increases the amount of bits that need to be exchanged, as formalized below.

\begin{proposition}[Theorem 6.1 in \cite{mittal_how_2026}]
    \label{prop:multi-layer-pointer-lower}
    For~$\ell = \cO(\sqrt{m}\,)$,
    \(
    \CC^R_{\nicefrac{\ell}{1000}}(\HistPC) = \Omega(m).
    \)
\end{proposition}

The deterministic case, which works even for~$\ell-1$ rounds, was established in~\cite{nisan_rounds_1991}. Mittal~\cite{mittal_how_2026} noted that it was possible to obtain the same asymptotic bound for the randomized case when restricted to~$\nicefrac{\ell}{1000}$ rounds instead.\footnote{For~$\ell-1$ rounds, there exists a randomized protocol using~$\Omega\left(\nicefrac{m}{\ell}\right)$ bits~\cite{rao_communication_2020}, with matching lower bound.}

Since, by~\Cref{thm:upper-bound-communication}, we know it is not possible to use~\Cref{thm:lower-no-round} for a non-trivial lower bound for~\kCore, we present a stronger version of the theorem, which considers lower bounds that include round restrictions, and which allows a graph \emph{function}~$F$, instead of simply a predicate~$P$.

A graph function~$F$ is a function mapping each graph to a string of bits. In the \CONGEST problem of computing a graph function~$F$ on network~$G$, each node has to output~$F(G)$ at the end of the algorithm. Note that a graph predicate is simply a graph function that outputs a single bit. We now slightly modify~\Cref{def:family} to allow graph functions.

\begin{definition}\label{def:family2}
    Let~$f: A \x B \to \zo$ be a function, and let $F$ be a graph function. A family of graphs~$\left\{G_{x,y} = (V, E_{x,y}) : (x, y) \in A \x B \right\}$
    with a partition~$V = V_A \cup V_B$, and possibly labels on the vertices and edges, is said to be a \emph{family of lower bound graphs w.r.t.~$f$ and~$F$} if the following properties hold:
    \begin{enumerate}
        \item only edges and labels in~$G[V_A]$ may depend on~$x$;
        \item only edges and labels in~$G[V_B]$ may depend on~$y$;
        \item $f(x,y)$ is uniquely determined by $F(G_{x, y})$.
    \end{enumerate}
\end{definition}

\begin{theorem}\label{thm:lower-rounds}
    Fix an integer~$r \geq 1$, a function~$f: A \x B \to \zo$, and a graph function~$F$. If there exists a family~$\{G_{x,y} = (V, E_{x,y})\}$ of lower bound graphs w.r.t.\ $f$ and~$F$, with~$V = V_A \cup V_B$ and $C = E[V_A, V_B]$, then every deterministic algorithm for computing~$F$ in the \CONGEST model requires~$\Omega(\min\{r, \;\CC_{r-1}(f)/|C|\cdot \log |V|\})$ rounds, and every randomized algorithm requires~$\Omega(\min\{r,\; \CC^R_{r-1}(f)/|C|\cdot \log |V|\})$ rounds.
\end{theorem}

\begin{proof}
    Let~$R(n)$ be the round complexity of a \CONGEST algorithm for graph function~$F$ in~$n$-node networks. Using the same simulation as for establishing~\Cref{thm:lower-no-round}, we obtain a communication protocol~$\pi$ for~$f$. This protocol runs in~$\cO(R(n))$ rounds by simulating each \CONGEST round with two alternating communication rounds. Moreover, it exchanges~${\cO(R(n) \cdot |C| \cdot \log n)}$ total bits as the players can only communicate $\cO(\log n)$ bits per round per edge in the cut.

    If~$\pi$ is deterministic, then either it takes at least~$r$ rounds, or it exchanges~$\Omega(\CC_{r-1}(f))$ bits. In the former case, we have $R(n) = \Omega(r)$. In the latter, $R(n) \cdot |C| \cdot \log n = \Omega(\CC_{r-1}(f))$. By combining both cases, we obtain~$R(n) = \Omega(\min\{r, \;\CC_{r-1}(f)/|C|\cdot \log n\})$. The case of a randomized~$\pi$ follows from the same arguments.
\end{proof}

\subsubsection{Using the Extended Framework}

We will start by deriving lower bounds for \MCVP, and later adapting them for~\kCore. We consider the variant of~\MCVP in \CONGEST in which each vertex must compute the value of the corresponding gate in the circuit.
The following proposition is the main technical reduction of this section.

\begin{proposition}\label{prop:mcvp-family}
    For any integers~$\ell, m \geq 2$, there exists a family of lower bound graphs ${\{G_{x,y} = (V, E_{x,y})\}}$ w.r.t.~\HistPC and~\MCVP, with~$V = V_A \cup V_B$ and~$C = E[V_A, V_B]$, such that ${|C| = \Theta(\ell \cdot \log m)}$,~$|V| = \Theta(\ell \cdot m \cdot \log (\ell \cdot m))$, with the graphs having diameter~$\cO(\log |V|)$.
\end{proposition}

This proposition means that, from a pointer chasing instance, we can obtain a graph which encodes a circuit such that, for each~$i \in [\ell]$, we can determine~$o_i$ based on the values of each of the gates. We first use the proposition to derive the lower bounds and then present its proof.

\begin{corollary}\label{cor:k-core-family}
    For any integers~$\ell, m \geq 2$, and~$k \geq 3$, there exists a family of lower bound graphs~$\{G_{x,y} = (V, E_{x,y})\}$ w.r.t.~\HistPC and~\kCore, with~$V=V_A \cup V_B$ and~$C = E[V_A, V_B]$, such that~$|C| = \Theta(\ell \cdot \log m)$, ${|V| = \Theta(\ell \cdot m \cdot \log (\ell \cdot m))}$, with the graphs having diameter~$\cO(\log |V|)$.
\end{corollary}

\begin{proof}
    For any input~$(x,y)$ for \HistPC, let $G_{x,y}$ be the graph given by~\Cref{prop:mcvp-family}. Then use the standard reduction from \MCVP to \kCore (see \cite[\S 3]{anderson_p-complete_1984}) to obtain $G'_{x,y}$. This reduction preserves the order of the number of vertices and cut sizes, and it is possible to recover the value of all gates of~$G_{x,y}$ from the $k$-core of $G'_{x,y}$. Therefore, $\{G'_{x,y}\}$ is a family that satisfies the statement.
\end{proof}

\begin{theorem}\label{thm:mcvp-round-lower}
    \MCVP on $n$-node $\cO(\log n)$-diameter networks requires $\Omega(\sqrt[3]{\nicefrac{n}{\log^3 n}})$ rounds for any randomized algorithm in the \CONGEST model.
\end{theorem}

\begin{proof}
    Suppose \MCVP can be solved in the \CONGEST model in $R(n)$ rounds.
    Consider an instance of \HistPC for some $\ell=\cO(\sqrt{m})$ that we will specify later.
    By~\Cref{prop:mcvp-family}, there exists some graph~$G = (V, E)$ with cut~$C = E[V_A, V_B]$, which encodes an instance of~\MCVP, such that, for each~$i \in [\ell]$, the value of~$o_i$ can be recovered from the values of the gates of the circuit, with~$|C| = \Theta(\ell \cdot \log m)$ and~$n=|V| = \Theta(\ell \cdot m \cdot \log(\ell \cdot m)) = \Theta(\ell \cdot m \cdot \log m)$.

    Using~\Cref{thm:lower-rounds} with $r = \nicefrac{\ell}{1000}+1$, and~\Cref{prop:multi-layer-pointer-lower}, we get that
    \[
        R(|V|) = \Omega\left(\min\left\{\frac{\ell}{1000}+1, \frac{\CC^R_{\nicefrac{\ell}{1000}}(\HistPC)}{\ell \cdot \log m \cdot \log |V|}\right\}\right) = \Omega\left(\min\left\{\ell, \frac{m}{\ell \cdot \log^2 m}\right\}\right),
    \]
    because $\ell = \cO(\sqrt{m})$. This is maximized when $\ell = m/(\ell \cdot \log^2 m)$ and therefore $\ell = \sqrt{m}/\log m = \cO(\sqrt{m})$, which implies that $|V| = \Theta(\sqrt{m^3}\,)$ and $R(|V|) = \Omega(\sqrt{m}/\log m)$. Changing variables,
    we obtain
    \[
        R(n) = \Omega(\sqrt[3]{n}/\log n) = \Omega\Big(\sqrt[3]{n/\log^3 n}\Big),
    \]
    using the fact that~$R(\cO(x)) = \cO(R(x))$ when~$R = \poly(x)$.
\end{proof}

\begin{theorem}\label{thm:k-core-lower}
    For any~$k \geq 3$,~\kCore on $n$-node $\cO(\log n)$-diameter networks requires $\Omega(\sqrt[3]{\nicefrac{n}{\log^3 n}})$ rounds for any randomized algorithm in the \CONGEST model.
\end{theorem}

\begin{proof}
    Identical to \Cref{thm:mcvp-round-lower}, but using \Cref{cor:k-core-family} instead of \Cref{prop:mcvp-family}.
\end{proof}

Note that the requirement that~$k \geq 3$ in the previous result is tight, in a sense, since~\kCore[2] can be solved in~$\cO(\log n)$ rounds on~$\cO(\log n)$-diameter networks, as we show in~\Cref{sec:two-core-upper}.

\begin{remark*}
    Theorem~\ref{thm:k-core-lower} also holds for any~$(2-\varepsilon)$-approximation of the core value of each vertex. It also holds for the directed~$k$-core variant, and the~$k$-truss problem. This is due to the known reductions from \MCVP to all of these problems~\cite[\S 3]{couto_hardness_2025}, and the fact that these reductions preserve the order of the cut sizes.
    In general, Theorem~\ref{thm:k-core-lower} holds for many \P-complete graph problems, and this technique for establishing hardness may be reused. This can be done for any problem~$\pi$ for which a reduction from~\MCVP exists where the size of the Alice-Bob cut in~\Cref{prop:mcvp-family} is roughly maintained, and that the values of all gates can be recovered. For example, it also follows for \LFMIS and \LFDVC.
\end{remark*}

With the motivation in place, we now prove this section's main technical result.

\begin{proof}[Proof of \Cref{prop:mcvp-family}]
    We will describe a family of lower bound graphs w.r.t.~\HistPC and \MCVP. At first, we consider circuits with unbounded diameter, fan-in, and fan-out.

    Consider an instance of~\HistPC. That is, we have functions $f_A, f_B: [m] \to [m]$. The input is given to two players, Alice and Bob, so that Alice knows~$f_A$, and Bob knows~$f_B$. To simplify notation, we take~$f_i$ to be $f_A$ if $i$ is odd, and $f_B$ if $i$ is even. We will build a monotone circuit that can be used to compute~$o_i = f_i(\cdots f_1(1) \cdots)$ for each $i \in [\ell]$.

    \begin{figure}[tb]
        \centering
        \input{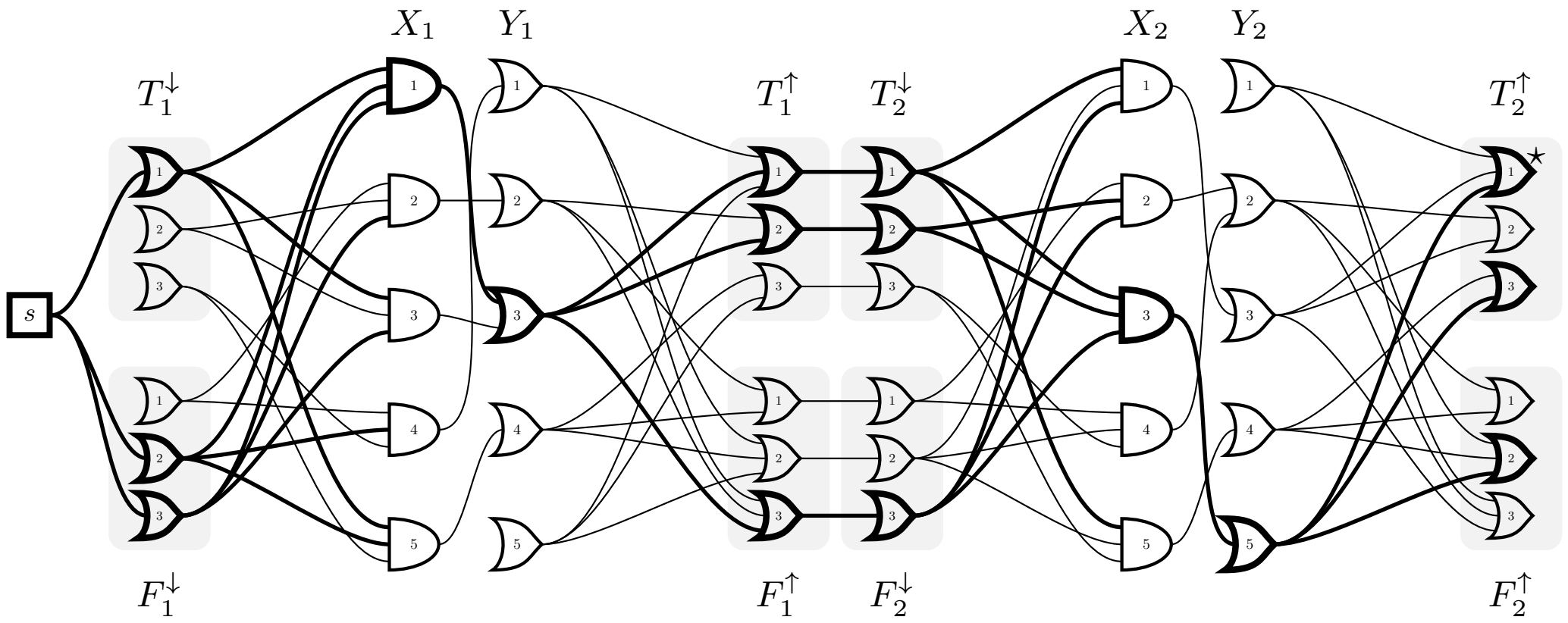}
        \caption{Example of the circuit created when~$\ell=2$,~$m=5$,~$f_A(i)=A_i$ and~$f_B(i)=B_i$ with~$A=[3,2,3,1,4]$ and~$B=[3,2,5,2,4]$. Gates and wires with value 1 are bold. In particular, the output gate (marked with a~$\star$) has value~$1$ because~$f_B(f_A(1)) = 5$.}
        \label{fig:mcvp-family-proof}
    \end{figure}

    \paragraph*{Construction of the circuit.}

    The circuit will have $\ell$ layers, where the~$i$th layer corresponds to function~$f_i$.
    Follow the description looking at the example in~\Cref{fig:mcvp-family-proof}.
    Layer~$i$ has AND-gates $X_i = \{\xx{i}{1}, \ldots, \xx{i}{m}\}$ and OR-gates $Y_i = \{\yy{i}{1}, \ldots, \yy{i}{m}\}$, as well as OR-gates
    \[
        \Fin{i} = \{\fin{i}{1}, \ldots, \fin{i}{\lgm}\} \;
        \mbox{and} \;
        \Fout{i} = \{\fout{i}{1}, \ldots, \fout{i}{\lgm}\},
    \]
    and
    \[
        \Tin{i} = \{\tin{i}{1}, \ldots, \tin{i}{\lgm}\}\;
        \mbox{and}\;
        \Tout{i} = \{\tout{i}{1}, \ldots, \tout{i}{\lgm}\}.
    \]
    Let $L_i = X_i \cup Y_i \cup \Fin{i} \cup \Fout{i} \cup \Tin{i} \cup \Tout{i}$ be the set of gates in layer~$i$.
    Now, let $j^k$ be the~$k$-th (least significant) bit of $j$ in its binary representation, and let
    \[
        {\Bin{i}{j} = \{\tin{i}{k}: j^k = 1\}} \cup \{\fin{i}{k}: j^k = 0\}.
    \]
    Define $\Bout{i}{j}$ similarly, but using $\tout{i}{k}$ and $\fout{i}{k}$. The wires between gates in layer $i$ are as follows, for each $j \in [m]$:
    \begin{itemize}
        \item from all gates in $\Bin{i}{j}$ to $\xx{i}{j}$,
        \item from $\xx{i}{j}$ to $\yy{i}{f_i(j)}$,
        \item from $\yy{i}{j}$ to all gates in $\Bout{i}{j}$.
    \end{itemize}
    Additionally, for each $i \in [\ell-1]$ and $k \in [\lgm]$, we have wires from $\fout{i}{k}$ to $\fin{(i+1)}{k}$ and from $\tout{i}{k}$ to $\tin{(i+1)}{k}$. Finally, add a 1-gate $s$ connected to $\Bin{1}{1}$. The output of the circuit is $\tout{\ell}{1}$.
    Alice will own $$V_A = \{s\}\cup \bigcup\limits_{i \text{ odd}} L_i,$$ while Bob will own $$V_B = \bigcup\limits_{i \text{ even}} L_i.$$ Note the wires in $E[V_A]$ depend only on the functions with odd indices (i.e., known by Alice), and the wires in $E[V_B]$ depend only on the functions with even indices (i.e., known by Bob). The wires in $E[V_A, V_B]$ are exactly those connecting $\Fout{i}, \Tout{i}$ to $\Fin{i+1}, \Tin{i+1}$, and do not depend on any function. Furthermore, the size of the cut $E[V_A, V_B]$ is $\cO(\ell \cdot \log m)$. One can think of getting rid of $\Fout{i}$ and $\Tout{i}$, connecting $y_{ij}$ directly to $\Fin{i+1}$, $\Tin{i+1}$, or~$X_{i+1}$, but this would increase the size of the $E[V_A, V_B]$ cut to $\Omega(m \cdot \ell)$.

    In order to bound the diameter of the circuit (when considered as an undirected graph), each player adds a 1-gate connected to all their AND-gates, and a 0-gate connected to all their OR-gates. Note this does not affect the value of the circuit, or the size of the cut, and it makes the diameter $\cO(1)$. To bound the fan-in and fan-out to 2, use the usual construction of replacing high-fan-in gates by balanced binary trees of gates of the same type, and high-fan-out gates by balanced binary trees of OR-gates. This increases the diameter by a logarithmic factor. The final circuit has $\cO(\ell \cdot m \cdot \log (\ell \cdot m))$ gates and wires, and diameter $\cO(\log (\ell \cdot m))$.

    \paragraph*{Correctness.}

    We can show by induction on $i$ that
    \begin{enumerate}
        \item $\xx{i}{j}$ evaluates to 1 if and only if $f_{i-1}(f_{i-2}(\cdots f_1(1) \cdots)) = j$,
        \item $\yy{i}{j}$ evaluates to 1 if and only if $f_i(f_{i-1}(\cdots f_1(1) \cdots)) = j$,
        \item gates in $\Fin{i}$ and $\Tin{i}$ evaluate to 1 if and only if they are in~$\Bin{i}{f_{i-1}(\cdots f_1(1) \cdots)}$, and
        \item gates in $\Fout{i}$ and~$\Tout{i}$ evaluate to 1 if and only if they are in~$\Bout{i}{f_i(f_{i-1}(\cdots f_1(1) \cdots))}$.
    \end{enumerate}
    When~$i=0$, (3) follows from our construction, because we added $s$ connected to $\Bin{1}{1}$ and no other gates in layer 1 have inputs. For $i > 0$, (3) holds because of (2) for layer~$i-1$, and the fact that~$\fin{i}{k}$ and~$\tin{i}{k}$ receive inputs only from~$\fout{(i-1)}{k}$ and~$\tout{(i-1)}{k}$, respectively. Now, (1) holds because of (3), and the fact that $\xx{i}{j}$ is connected to $\Bin{i}{j}$ and any~$k \neq j$ is connected to some gate in $\Bin{i}{k}$ that evaluates to 0. Moreover, (2) holds because of (1) and the fact that $\yy{i}{j}$ is connected to all $\xx{i}{k}$ such that $f_i(k) = j$. Finally, (4) holds because of (2) and the connections from $\yy{i}{j}$ to $\Bout{i}{j}$.

    Thus, we can correctly determine~$o_i = j$ where~$\yy{i}{j}$ is the only gate on that column with value 1. Hence, we have a family of lower bound graphs w.r.t.~\HistPC and~\MCVP, where each graph is the undirected version of the circuit, and each vertex has information (labels) about which type of gate it is and the orientation of the edges incident on it.
\end{proof}

\section{An Efficient Algorithm for Computing the 2-Core}
\label{sec:two-core-upper}

In this section, we present algorithms for degeneracy in \CONGEST. Note that from an algorithm for~\kCore we can easily create one for~$\kDeg[(k-1)]$.\footnote{If any vertex is in the $k$-core, then the graph is not~$(k-1)$-degenerate, so this can be computed in additional~$\cO(D)$ rounds.} Therefore, for the remainder of the section, we focus on algorithms for the~$k$-core problem.

For~$k\geq 3$, the straightforward~$\cO(n)$-round algorithm matches the lower bound of~\Cref{cor:3core-lower-cvp} up to subpolynomial factors. However, for~$k=2$, we provide a more efficient algorithm that runs in~$\cO(D)$ rounds on graphs of diameter~$D$, thus matching the known~$\Omega(D)$ lower bound, which holds even for the \local model~(see~\cite[\S1]{chan_distributed_2019}).

We start by describing a distributed \CONGEST algorithm proposed in~\cite{chan_distributed_2019} for computing the core values of a given graph $G$.
During the execution, each node~$u$ of $G$ maintains a local estimate of its core value, which is initialized to its degree.
Throughout the execution, its estimate is at least its true core value.
At each round, each node $u$ broadcasts its core value estimate to its neighbors.
Once $u$ receives all current estimates from its neighbors, it updates its estimate to the largest $x$ such that $u$ has at least $x$ neighbors with estimate at least $x$. For an $n$-node graph, after $n$ rounds, the estimates are the true core values~\cite[\S I.A]{chan_distributed_2019}, so all nodes terminate.

\begin{remark*}
    Even on $n$-node graphs with constant diameter, the number of rounds needed for the algorithm to stabilize (i.e., every node has its true core value) might be $\Theta(n)$.
    Indeed, for the graph consisting of an $(n-1)$-node path plus a universal node, the algorithm takes $\left\lfloor \frac{n-2}{2}\right\rfloor$ rounds to stabilize. In fact, at the end of round $i\in\{1,2,\ldots,\left\lfloor \frac{n-4}{2}\right\rfloor\}$, every node at distance at most $i$ from the endpoints of the $(n-1)$-node path has set its core value estimate to $2$, while all other nodes, including the universal node, have their estimate equal to $3$. Finally, in round $\left\lfloor \frac{n-2}{2}\right\rfloor$, every node sets its core value estimate to its true core value, which is $2$.
\end{remark*}

We now present a more efficient \CONGEST algorithm when~$k=2$.

\begin{theorem}
    \kCore[2] can be solved in $\cO(D)$ rounds in the \CONGEST model on $n$-node networks of diameter~$D$.
\end{theorem}

\begin{proof}
    Let $G$ be a graph of diameter $D$.
    The first phase of our algorithm starts with the execution of the standard  procedure for selecting as leader a node~$s$ of $G$ and computing a breadth-first search tree~$T$ of $G$ rooted at $s$. This takes $\cO(D)$ rounds in the \CONGEST model (see~\cite[\S5.5-5.6]{distributed_algorithms_2020} for details).
    The execution of this procedure determines, for each node~$v$ of~$G$, its distance to the root $s$, a pointer $\prnt[v]$ to its parent, and the collection of identifiers of~$v$'s children in $T$.
    In the next phases, each node $v$ uses two additional Boolean variables, $\mkup[v]$ and $\mkdown[v]$, to identify $v$ as unmarked, marked up, marked down, or marked both up and down. Initially, these two variables are set to $\false$ at each node.

    At the start of the second (upcast) phase of our algorithm,
    each node $v$ incident to a non-tree edge sets variable $\mkup[v]$ to~$\true$ conveying that it belongs to a cycle in~$G$.
    Afterward, every leaf~$v$ of~$T$ upcasts $\mkup[v]$ to its parent.
    When an internal node $u$ of~$T$ has seen all messages being upcast by its children, it sets $\mkup[u]$ to $\true$ if it received at least one
    $\true$ value. The second phase finishes when the root of $T$ has processed all its children's messages.
    Besides this, in preparation for the third phase, during the upcast phase, $v$ sets $\mkdown[v]$ to~$\true$ if it received at least two $\true$ values from its children.

    During the third (downcast) phase, messages are propagated downward through $T$ starting from the root $s$, which broadcasts $\mkdown[s]$.
    When a node~$v$ receives the message from its parent, it sets
    $\mkdown[v]$ to~$\true$ if the message is~$\true$, and then it propagates $\mkdown[v]$ to its children.
    See \Cref{fig:2core} for an example.

    Finally, a node $v$ claims that it belongs to the $2$-core of $G$ if and only if both $\mkup[v]$ and $\mkdown[v]$ equal~$\true$ at the end of the downcast phase.

    \begin{figure}[ht]
        \centering
        \begin{tikzpicture}[scale=0.9,
    hollow/.style={circle, draw, thick, minimum size=0.5cm, inner sep=0pt, font=\sffamily\bfseries},
    in_2core/.style={hollow, fill=blue!30},
    not_in_2core/.style={hollow, fill=gray!30},
    tree_edge/.style={thick},
    dashed_edge/.style={thick, dashed, red!80},
    arrowmarkstyle/.style={font=\tiny, inner sep=1pt},
    upmark/.style={arrowmarkstyle, blue, ->},
    downmark/.style={arrowmarkstyle, purple, ->}
]
    \node [not_in_2core] (A) at (-1, 1) {A};
    \node [not_in_2core] (B) at (-2.2, 0) {B};
    \node [in_2core] (C) at (-1, 0) {C};
    \node [in_2core] (D) at (-2.1, -1) {D};
    \node [in_2core] (E) at (0.1, -1) {E};
    \node [in_2core] (F) at (-2.6, -2) {F};
    \node [in_2core] (G) at (-1.6, -2) {G};
    \node [in_2core] (H) at (-2.6, -3) {H};
    \node [not_in_2core] (I) at (-0.4, -2) {I};
    \node [in_2core] (J) at (0.6, -2) {J};
    \node [in_2core] (K) at (0.1, -3) {K};
    \node [in_2core] (L) at (1.1, -3) {L};

    \draw [tree_edge] (A) -- (B);
    \draw [tree_edge] (A) -- (C);
    \draw [tree_edge] (C) -- (D);
    \draw [tree_edge] (C) -- (E);
    \draw [tree_edge] (D) -- (F);
    \draw [tree_edge] (D) -- (G);
    \draw [tree_edge] (F) -- (H);
    \draw [tree_edge] (E) -- (I);
    \draw [tree_edge] (E) -- (J);
    \draw [tree_edge] (J) -- (K);
    \draw [tree_edge] (J) -- (L);

    \draw [dashed_edge] (G) to (H);
    \draw [dashed_edge] (K) to (L);

    \draw[upmark] (A.east) ++(0.1, 0.08) -- ++(0, 0.2);
    \draw[upmark] (C.east) ++(0.1, 0.08) -- ++(0, 0.2);
    \draw[downmark] (C.east) ++(0.1, -0.08) -- ++(0, -0.2);
    \draw[upmark] (D.west) ++(-0.1, 0.08) -- ++(0, 0.2);
    \draw[downmark] (D.west) ++(-0.1, -0.08) -- ++(0, -0.2);
    \draw[upmark] (E.east) ++(0.1, 0.08) -- ++(0, 0.2);
    \draw[downmark] (E.east) ++(0.1, -0.08) -- ++(0, -0.2);
    \draw[upmark] (F.west) ++(-0.1, 0.08) -- ++(0, 0.2);
    \draw[downmark] (F.west) ++(-0.1, -0.08) -- ++(0, -0.2);
    \draw[upmark] (G.east) ++(0.1, 0.08) -- ++(0, 0.2);
    \draw[downmark] (G.east) ++(0.1, -0.08) -- ++(0, -0.2);
    \draw[upmark] (H.west) ++(-0.1, 0.08) -- ++(0, 0.2);
    \draw[downmark] (H.west) ++(-0.1, -0.08) -- ++(0, -0.2);
    \draw[downmark] (I.west) ++(-0.1, -0.08) -- ++(0, -0.2);
    \draw[upmark] (J.east) ++(0.1, 0.08) -- ++(0, 0.2);
    \draw[downmark] (J.east) ++(0.1, -0.08) -- ++(0, -0.2);
    \draw[upmark] (K.west) ++(-0.1, 0.08) -- ++(0, 0.2);
    \draw[downmark] (K.west) ++(-0.1, -0.08) -- ++(0, -0.2);
    \draw[upmark] (L.east) ++(0.1, 0.08) -- ++(0, 0.2);
    \draw[downmark] (L.east) ++(0.1, -0.08) -- ++(0, -0.2);
\end{tikzpicture}
        \caption{
            The final state of the 2-core algorithm on an example graph.
            The solid black edges form a BFS tree rooted at node {\bf \textsf{A}}.
            The dashed red edges are non-tree edges.
            The small arrows next to a node~$v$ indicate its final state: a blue upward arrow for $\mkup[v]$ and a purple downward arrow for $\mkdown[v]$.
            A node is in the 2-core, colored blue in the image, if and only if it has both marks at the end of the algorithm.
        }
        \label{fig:2core}
    \end{figure}
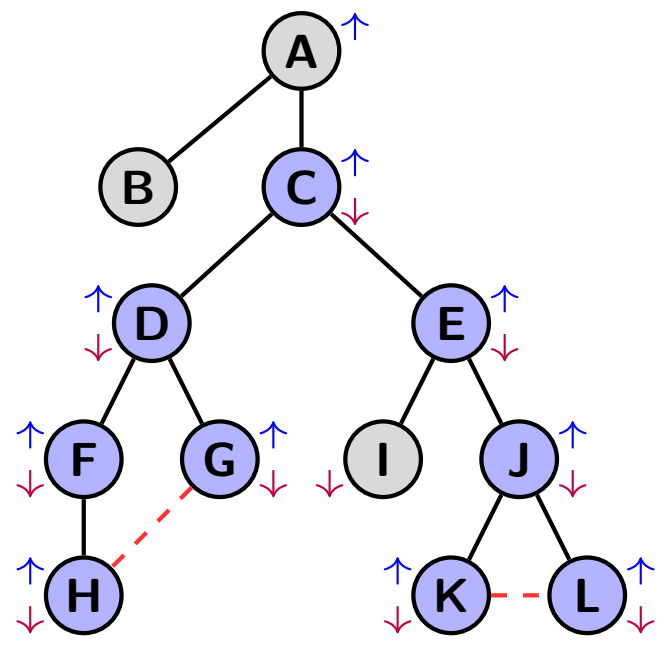

    By construction, our algorithm completes in $\cO(D)$ rounds in \CONGEST, since building the breadth-first search tree and both the upcast and downcast phases take $\cO(D)$ rounds, as the tree has depth $\cO(D)$.

    The following lemmas refer to the values of $\mkup$ and $\mkdown$ at the end of the execution of the algorithm. We denote by $T_v$ the subtree of $T$ rooted at~$v$.

    \begin{lemma}\label{lem:up}
        $\mkup[v] = \true$ if and only if a node in $T_v$ is in the 2-core of $G$.
    \end{lemma}
    \begin{proof}\renewcommand{\qedsymbol}{\diamo}
        If $\mkup[v] = \true$, then there is a descendant (in $T$) of $v$ incident to a non-tree edge. This descendant, which is a node of $T_v$, is in a cycle in $G$, and thus is in the 2-core of $G$. On the other hand, if $v$ is in the 2-core of $G$, then let $u$ be a descendant of $v$ in the 2-core with maximum depth in $T$.
        Each of the at least two edges in the 2-core incident to~$u$ does not go to a descendant of $u$. Therefore, at least one of these edges is not in $T$, and hence $\mkup[v] = \mkup[u] = \true$.
    \end{proof}

    \begin{lemma}
        $\mkup[v] = \mkdown[v] = \true$ if and only if $v$ is in the $2$-core of $G$.
    \end{lemma}
    \begin{proof}\renewcommand{\qedsymbol}{\diamo}
        The proof is by induction on the depth of $v$ in $T$.
        First, assume $\mkup[v] = \mkdown[v] = \true$.
        If $v \neq s$ and $\mkdown[\prnt[v]] = \true$, then $\mkup[\prnt[v]]=\true$ because $\mkup[v]=\true$, and, by induction, $\prnt[v]$ is in the 2-core of $G$. By \Cref{lem:up}, there is also a path in $T_v$ from $v$ to a node in the $2$-core. So, there are two edge-disjoint paths from $v$ to vertices in the 2-core, which implies that~$v$ is also in the 2-core.
        If~$v = s$ or $\mkdown[\prnt[v]] = \false$, then, as $\mkdown[v]=\true$, $v$ must have two children $u$ and $w$ such that $\mkup[u]=\mkup[w]=\true$. By \Cref{lem:up}, there are two edge-disjoint paths from $v$ to vertices in the 2-core. So, $v$ is in the $2$-core of~$G$.

        Now, assume $v$ is in the 2-core of $G$. By \Cref{lem:up}, $\mkup[v] = \true$.
        If $v \neq s$ and $\prnt[v]$ is in the 2-core, then, by induction, $\mkdown[\prnt[v]] = \true$, and hence $\mkdown[v] = \true$.
        If $v=s$ or $\prnt[v]$ is not in the 2-core, then no non-tree edge is incident to $v$, because in a BFS tree the fundamental cycle of a non-tree edge~$vu$ always includes $\prnt[v]$.
        There are at least two edges $vu$ and $vw$ in the 2-core, and~$u$ and $w$ must be distinct children of~$v$, since $\prnt[v]$ is not in the 2-core.
        By \Cref{lem:up}, $\mkup[u] = \mkup[w] = \true$, thus $\mkdown[v] = \true$.
    \end{proof}

    This finishes the proof of correctness of the algorithm.
\end{proof}

We have thus shown precisely the threshold of hardness of degeneracy problems in~\CONGEST. While \kCore[2] (and \kDeg[1]) can be solved efficiently (i.e., they need $\cO(D)$ rounds), \kCore[3] (and \kDeg[2]) cannot (i.e., they need $\widetilde{\Omega}(D + \sqrt{n})$ rounds).

\section{Conclusion}

We provided a series of lower bounds on the round complexity of computing the degeneracy of a network in \CONGEST.
We established unconditional lower bounds showing that, for any $k \geq 2$, solving~\kDeg requires $\widetilde{\Omega}(D+\sqrt{n})$ rounds, even with a randomized algorithm, and even for $(2-\varepsilon)$-approximation.
Furthermore, we provided conditional lower bounds suggesting that the problem is likely even harder.
Specifically, we showed that algorithms running in $\cO(n^{1-\varepsilon})$ rounds would imply significant breakthroughs in complexity theory, such as the circuit value problem admitting a polynomial parallel speedup, unless they consume high local computation time.
These results provide strong evidence that computing the degeneracy requires $\widetilde{\Omega}(n)$ rounds, even in networks of constant maximum degree and logarithmic diameter.

An interesting open problem is to close the gap between our unconditional polynomial lower bounds and the straightforward linear upper bound.
Another possible path forward is to generalize the lower bounds. We note in Section~\ref{subsubsec:other-P-complete} that our lower bounds hold for other graph problems, such as \LFMIS, {\LFDVC}, and \MaxFlow. Properly formalizing the subset of~{\P}-complete problems for which our conditional or unconditional lower bounds hold would make clear which results from~{\P}-completeness translate (totally or partially) to the \mbox{\CONGEST} model, allowing us to reuse established theoretical results from complexity theory instead of having to prove them all over again in the  \CONGEST model. Similar lower bounds for {\P}-complete problems have also emerged from other models, such as the semi-streaming model~\cite{assadi_polynomial_2019} and the dynamic graph model~\cite{couto_hardness_2025}, and this generalization could benefit those areas as well.

\bibliographystyle{splncs04}
\bibliography{references}


\end{document}